\documentclass[journal,oneside]{IEEEtran}
\usepackage{amsmath,amssymb }

\usepackage{cite}

\usepackage{acronym}

\usepackage{tikz}
\usetikzlibrary{shapes,arrows}
\usetikzlibrary{shapes.geometric}

\usepackage{pgfplots}
\usepackage{pgfplotstable}
\usepgfplotslibrary{colorbrewer}
\usepgfplotslibrary{groupplots}

\pgfplotsset{compat=1.17}

\usepackage{algorithm}%
\usepackage{algpseudocode}%
\makeatletter
\algnewcommand{\LineComment}[1]{\Statex{\hskip\ALG@tlm \(\triangleright\) \textit{#1}}}
\makeatother

\pgfplotsset{ 
    snrgap/.style={
        legend cell align={left},
        legend style={font=\footnotesize,nodes={scale=0.8,transform shape}},
        minor tick num=1,
        tick label style={font=\small},
        ylabel style={yshift=-0.1cm},
        xlabel={SNR [dB]},
        ylabel={Gap to AWGN C [bit/2Dsym]},
        every axis plot/.append style={thick},
        grid=both,
        minor grid style={dotted},
    },
    legend image with text/.style={
        legend image code/.code={%
          \node[anchor=center] at (0.3cm,0cm) {#1};
        }
    },
    colormap={isoL-10}{rgb255=(232,57,229) rgb255=(155,75,252) rgb255=(90,100,252) rgb255=(26,137,195) rgb255=(10,153,115) rgb255=(9,157,19) rgb255=(34,155,3) rgb255=(123,141,2) rgb255=(191,112,12) rgb255=(255,58,41)},
    colormap={isoL-10-R}{ rgb255=(255,58,41) rgb255=(191,112,12)  rgb255=(123,141,2) rgb255=(34,155,3) rgb255=(9,157,19) rgb255=(10,153,115) rgb255=(26,137,195) rgb255=(90,100,252) rgb255=(155,75,252) rgb255=(232,57,229)},
    colormap={isoL-20}{rgb255=(232,57,229) rgb255=(194,65,244) rgb255=(158,74,252) rgb255=(126,85,254) rgb255=(96,97,254) rgb255=(65,113,238) rgb255=(35,131,207) rgb255=(14,146,171) rgb255=(10,151,135) rgb255=(10,154,88) rgb255=(9,157,40) rgb255=(9,158,8) rgb255=(15,157,4) rgb255=(46,153,3) rgb255=(90,147,3) rgb255=(131,140,2) rgb255=(164,128,5) rgb255=(195,110,13) rgb255=(225,86,26) rgb255=(255,58,41)},
}

\tikzstyle{startstop} = [rectangle, rounded corners, minimum width=2cm, minimum height=0.75cm,text centered, draw=black, fill=red!30]
\tikzstyle{io} = [trapezium, trapezium left angle=70, trapezium right angle=110, minimum width=2cm, minimum height=0.75cm, text centered, draw=black, fill=blue!30]
\tikzstyle{process} = [rectangle, minimum width=2cm, minimum height=0.75cm, text centered, draw=black, fill=orange!30]
\tikzstyle{decision} = [kite, kite vertex angles=135, inner sep=0pt, minimum width=2cm, minimum height=0.75cm, text centered, draw=black, fill=green!30]
\tikzstyle{arrow} = [thick,->,>=stealth]
 
\acrodef{ASE}{amplified spontaneous emission}
\acrodef{APSK}{amplitude and phase-shift keyed} 
\acrodef{AIR}{achievable information rate}
\acrodef{BER}{bit-error rate}
\acrodef{BFGS}{Broyden Fletcher Goldfarb Shanno}
\acrodef{BICM}{bit interleaved coded modulation}
\acrodef{AWGN}{additive white Gaussian noise}
\acrodef{BRGC}{binary reflected Gray code}
\acrodef{CDF}{cumulative distribution function}
\acrodef{CM}{coded modulation}
\acrodef{FEC}{forward error correction}
\acrodef{GHQ}{Gauss-Hermite quadrature}
\acrodef{GMI}{generalised mutual information}
\acrodef{GS}{geometrical shaping}
\acrodef{GS-2D}{geometrically shaped 2-dimensional}
\acrodef{GS-4D}{geometrically shaped 4-dimensional}
\acrodef{LDPC}{low-density parity check}
\acrodef{MI}{mutual information}
\acrodef{OS}{orthant symmetric}
\acrodef{PS}{probabilistic shaping}
\acrodef{SNR}{signal-to-noise ratio}
\acrodef{SR1}{symmetric rank-one}
\acrodef{SSFM}{split-step Fourier method}
\acrodef{QAM}{quadrature amplitude modulation}
\acrodef{QPSK}{quadrature phase shift keyed}

\newcommand{\E}{\int_{\mathbb{R}^{2N}} f(\vect{z})}
\newcommand{\dz}{\mathrm{d}\vect{z}}
\newcommand{\mat}[1]{\ensuremath{\boldsymbol{#1}}}
\newcommand{\ddx}[1]{\ensuremath{\frac{\partial}{\partial #1}}}

\newcommand{\vect}[1]{\ensuremath{{\underline{\smash{#1}}}}}

\newcommand{\setJi}{\ensuremath{\widetilde{\mathcal{J}_i}}}
\newcommand{\setJn}{\ensuremath{\widetilde{\mathcal{J}_{n}}}}
\newcommand{\setIbk}{\ensuremath{\mathcal{I}_k^b}}
\newcommand{\settIbk}{\ensuremath{\widetilde{\mathcal{I}_n}_k^b}}

\pgfplotsset{%
  log x ticks with fixed point/.style={%
      xticklabel={%
        \pgfkeys{/pgf/fpu=true}
        \pgfmathparse{exp(\tick)}%
        \pgfmathprintnumber[fixed relative, precision=3]{\pgfmathresult}
        \pgfkeys{/pgf/fpu=false}
      }
  },
}

\tikzset{%
    square/.style={regular polygon,regular polygon sides=4,inner sep=0,draw},
    qamlabel/.style={fill=white, font=\footnotesize, anchor=south, inner sep=1pt},
}

\begin{document}
\title{High-Cardinality Geometrical Constellation Shaping for the Nonlinear Fibre Channel}

\author{Eric~Sillekens,~\IEEEmembership{Member,~IEEE,}
        Gabriele~Liga,~\IEEEmembership{Member,~IEEE,}
        Domani\c{c}~Lavery,~\IEEEmembership{Member,~IEEE,}
        Polina Bayvel,~\IEEEmembership{Fellow,~IEEE,}
        and~Robert.~I.~Killey,~\IEEEmembership{Fellow,~IEEE}%
\thanks{This work was supported in part by a UK EPSRC
programme under Grant TRANSNET EP/R035342/1. The work of G.~Liga has received funding by the EuroTechPostdoc programme under the European Union’s Horizon 2020 research and innovation programme (Marie Skłodowska-Curie grant agreement No. 754462). Domani\c{c} Lavery was supported by the Royal Academy of Engineering under the Research Fellowships scheme}%
\thanks{Eric Sillekens, Polina Bayvel, and Robert I. Killey are with the Optical Networks Group, Department of Electronic and Electrical Engineering, University College London, WC1E 7JE London, U.K.}%
\thanks{Domani\c{c} Lavery was with the Optical Networks Group, Department of Electronic and Electrical Engineering, University College London, WC1E 7JE London, U.K.. He is now with Infinera Canada Inc., Ottawa, ON, Canada}%
\thanks{Gabriele Liga is with the Signal Processing Systems (SPS) Group, Department of Electrical Engineering, Eindhoven University of Technology, 5600 MB Eindhoven, The Netherlands}%
}

\markboth{High-Cardinality Geometrical Constellation Shaping for the Nonlinear Fibre Channel \today}{High-Cardinality Geometrical Constellation Shaping for the Nonlinear Fibre Channel \today}

\maketitle

\begin{abstract}
This paper presents design methods for highly efficient optimisation of geometrically shaped constellations to maximise data throughput in optical communications. 
It describes methods to analytically calculate the information-theoretical loss and the gradient of this loss as a function of the input constellation shape. The gradients of the mutual information (MI) and generalised mutual information (GMI) are critical to the optimisation of geometrically-shaped constellations.
The analytically derived gradients of the achievable information rate metrics with respect to the input constellation are presented. The proposed method allows for improved design of higher cardinality and higher-dimensional constellations for optimising both linear and nonlinear fibre transmission throughput. Near-capacity achieving constellations with up to 8192 points for both 2 and 4 dimensions are presented. In the best case, a GMI value within 0.06 bit/2Dsymbol of the additive white Gaussian noise channel (AWGN) capacity was achieved. Additionally, a design algorithm reducing the design computation time from days to minutes is introduced, allowing for the design of optimised constellations for both linear AWGN and nonlinear fibre channels over a wide range of signal-to-noise ratio values.
\end{abstract}
\IEEEpeerreviewmaketitle

\section{Introduction}

There is renewed interest in the use of capacity-approaching constellation shaping to further increase the throughput and reach of optical fibre communication systems while providing finer granularity flexible rates.
Constellation shaping can be divided into two flavours: \ac{PS} and \ac{GS}. In \ac{PS}, the constellation points are transmitted with unequal probabilities in order to maximise the \ac{MI} in a given channel, while in \ac{GS} we keep the constellation probabilities uniform and instead change the position of the equiprobable constellation points.

Recently, constellation shaping has contributed to the increase of the system information spectral density in optical fibre communication systems, where the constellations were designed for the linear \ac{AWGN} channel. 
There have been noteworthy results achieved employing constellation shaping in the optical fibre channel using \ac{PS} \cite{Ghazisaeidi2016Transoceanic,Olsson2018Probabilistically}, \ac{GS} \cite{Ionescu2019Transmission,Galdino2020Optical,Wakayama2019Increasing} and hybrid combinations of both PS and GS \cite{Cai2018Transmission}.
To maximise the benefits from shaping, the cardinality can be increased to have a greater impact on system performance. When the optimum is not known, the numerical optimisation of the constellation shape to maximise the information rate in a given channel becomes rapidly infeasible as constellation size and dimensionality increase. Larger constellations allow for more degrees of freedom when shaping a constellation for the channel, but that comes at the expense of increased computational complexity. To address this, machine learning has been moderately successful in the design of \ac{GS} constellations \cite{Jones2018Deep,Essiambre2020Increasing,Veeru2020End}, to minimise information-theoretical loss, and maximise potential data throughput. In these papers, the machine learning framework is used to obtain an objective function for the increase of performance of the constellation being designed, for which a gradient descent strategy is then employed.

Multiple approaches have been considered. Amongst them is a gradient descent for method the design of tailored probability mass functions for \ac{PS} constellations for fibre nonlinearity tolerance \cite{Fehenberger2016Probabilistic,Geller2016Shaping,Renner2017Experimental,Sillekens2018Simple}. Similarly, \ac{GS} constellations have been designed for fibre nonlinearity \cite{Zhang2017Design,Sillekens2018Experimental,chen2020analysis} or laser phase noise tolerance \cite{Dzieciol2020Geometric}. However, in general, the optimum geometric constellation shape for such a channel is still not known.

In the case of \ac{GS} constellation design, the complexity dramatically increases with constellation order. Both the evaluation of the \ac{MI} function and the calculation of the gradient function scale poorly with the number of constellation points, making the optimisation of multi-dimensional and/or high-order constellations prohibitively time-consuming. A contributing factor in this is the requirement to compute the gradient of the \ac{MI} or \ac{GMI} using exploratory steps, i.e., one \ac{MI} or \ac{GMI} calculation for each constellation point; an already costly calculation whose complexity scales rapidly with the number of constellation points. Recent works have proposed the use of orthant symmetry\cite{chen2020analysis}, where only a limited subset of the constellation points are optimised in a single orthant, and the remainder are formed as reflections in the other orthants. Although this method greatly simplifies the problem, the design of a constellation within a single orthant still requires a computationally expensive step for every constellation point within the orthant and, therefore, will ultimately severely constrain the maximum order of the constellations that can be designed.
 
This paper presents a method to analytically calculate the information-theoretical loss and the gradient of this loss as a function of the input constellation shape. The gradients of the \ac{MI} and \ac{GMI} are critical to the optimisation of geometrically-shaped constellations. In the process of numerical optimisation, the gradient provides a direction in which to increase or decrease the objective function. While the resulting constellations are not guaranteed to be optimal, significant improvements can be obtained.

Usually, the gradients are calculated using a finite difference method, where for every constellation point the cost function is re-evaluated with a small but finite difference to estimate the partial derivative. However, this method scales poorly with an increasing number of constellation points, resulting in an infeasibly complex task for large cardinality constellations, e.g., constellations with more than 1024 points. A more efficient method is to use automatic differentiation \cite{Beda1959Pfa,Rall1981Automatic}. Automatic differentiation offers faster numerical gradient calculation than the finite difference methods, albeit in general not as fast as calculating the gradient analytically. In this paper, the loss gradient is calculated analytically, an approach that has only been used for \ac{BER} \cite{Foschini1974Optimization} and has, to date, never been applied in optical communications and never used for \ac{GMI}.
Using the analytical expressions for these gradients speeds up the optimisation process by a factor corresponding to the constellation cardinality, i.e., only one calculation is needed for the whole constellation instead of a calculation for each constellation point separately, and is, therefore, an invaluable method for constellation design.

 The remainder of the paper is organised as follows. In Section~\ref{sec:preliminaries}, the optimisation problem is introduced and, in Section~\ref{sec:gradient_descent}, the optimisation method used in this work is shown. In Section~\ref{sec:derivative}, an efficient gradient calculation in introduced. Then, in Section~\ref{sec:chain_rule}, the gradient is extended to account for the nonlinearity of the fibre channel. After which, in Sections~\ref{sec:results} and~\ref{sec:resultnonlin}, we present results obtained using the algorithms shown in this work for the \ac{AWGN} channel and the nonlinear fibre channel, respectively, with the conclusions in Section~\ref{sec:conclusions}.

\section{Preliminaries}
\label{sec:preliminaries}

This section consists of four subsections: A introduces the notation used throughout the paper;  B describes the channels used; C describes the achievable information rates used in the paper for quantifying the performance of the designed constellations. Finally, D describes the optimisation algorithm based on gradients for nonlinear optimisation problems.

\subsection{Notation}

In this work, we use the following notation. Let $\mathbb{R}$ and $\mathbb{C}$ denote the real and complex numbers respectively. For random variables we use uppercase letters, e.g., $X$. We use lowercase letters (e.g.~$x$) and underlined lowercase letters (e.g.~$\vect{x}$) for ordinary scalar and vector variables, respectively. Matrices are denoted by boldface letters, e.g. $\mat{x}$ or $\mat{I}$. In particular, the matrix containing the $M$ possible constellation symbols coordinates is denoted as $\mat{x}=[\vect{x}_1,\ldots,\vect{x}_M]^T$ where the constellation symbol $\vect{x_i}=[x_{i,1},x_{i,2},\ldots,x_{i,2N}]$ represents the location of the $i$-th point in $2N$ real dimensions, where $N$ represents the number of phase and amplitude pairs.
To denote a set we use a calligraphic letter, $\mathcal{J}=\{1,\ldots,M\}$ is the set of all $M$ indices and $\setJn= \mathcal{J}\setminus\{n\}=\{j\in \mathcal{J}; j\neq n\}$ is the set of all indices excluding $n$. To denote a subset of constellation points we use the boldface subscripted notation $\mat{x}_\mathcal{S} = [\vect{x}_{i_1},\ldots,\vect{x}_{i_Q}]^T$ where $\mathcal{S} = \{i_1,\ldots,i_Q\} $ is the set of indices selecting the constellation symbols. 
Let $||\mat{x}||_F = \sqrt{\sum_{i=1}^M\sum_{j=1}^{2N}||x_{i,j}||^2}$ be the Frobenius norm of matrix $\mat{x}\triangleq \{x_{i,j}\}$ \cite[p.~55]{Golub1996Matrix}. 
For the probability density functions (pdfs), the conventional notation is used, where, for instance, $f_X(\vect{x})$ denotes the pdf of the random variable $X$ and $f_{Y|Y}(\vect{y}|\vect{x})$ denotes the conditional pdf of the random variable $Y$ given $X$. Let $|\cdot|$ be the absolute value of a scalar. Let $||\cdot||^2$ and $\langle\cdot,\cdot \rangle$ be the squared vector norm and inner product respectively.  Let $(\cdot)^H$ be the conjugate transpose, also known as the Hermitian transpose.  

The paper uses matrix calculus, meaning that partial derivatives of a multivariate function are organised in a Jacobian matrix or tensor. Let $\mat{f}(\mat{x}):\mathbb{R}^{M\times2N}\rightarrow\mathbb{R}^{N\times2N}$ be a function that maps matrix $\mat{x}$ to another matrix.
For this function, the Jacobian in tensor form is defined as:
\begin{align}
    \mat{J}_{\mat{f}}(\mat{x}) = \begin{bmatrix}
    \dfrac{\partial \mat{f}}{\partial \vect{x}_1} & \cdots & \dfrac{\partial \mat{f}}{\partial \vect{x}_n} \end{bmatrix}
\end{align}
\begin{equation}
    (\mat{J}_{\mat{f}}(\mat{x}))_{i,j,k,l} = \dfrac{\partial f_{i,j}}{\partial x_{k,l}}
\end{equation}
where $\dfrac{\partial f_{i,j}}{\partial x_{k,l}}$ is the partial derivative of the $i,j$-th output to the $k,l$-th input.

When function $f(\vect{x}):\mathbb{R}^M\rightarrow\mathbb{R}$ maps to a real scalar, then the gradient is defined as $\nabla f(\vect{x})=\mat{J}^H_\vect{f}(\vect{x})$. Let $(\vect{f}\circ\vect{g})(\vect{x})= \vect{f}(\vect{g}(\vect{x}))$ be a composition of two functions $\vect{f}(\vect{x})$ and $\vect{g}(\vect{x})$. Then, the Jacobian is defined as $\mat{J}_{\vect{f}\circ\vect{g}}(\vect{x}) = \mat{J}_\vect{f}(\vect{g}(\vect{x}))\mat{J}_\vect{g}(\vect{x})$. If function $\vect{g}(\vect{x})$ maps a vector to a vector, the Jacobian is a matrix. If the function maps a matrix to a matrix, the Jacobian is a tensor.

\subsection{Channel Models}\label{sec:channel_model}

In this paper, we study the optimisation of MI and GMI in two different channels: i) an \ac{AWGN} channel, and ii) a discrete-time optical fibre channel. The \ac{AWGN} channel is modelled as
\begin{align}
    f_{Y|X}(\vect{y}|\vect{x}) &= \frac{1}{(\pi \sigma_z^2)^N}\exp\left(\frac{-||\vect{y} -\vect{x}||^2}{\sigma_z^2}\right)  
\end{align}
with noise variance $\sigma_z^2$ per $2$ real dimensions. Variable $X$ denotes the transmitted symbols, Gaussian noise is added and then variable $Y$ denotes the received symbols; $\vect{x}$ and $\vect{y}$ are instances of those variables; $f_X(\vect{x})$ and $f_Y(\vect{y})$ are the corresponding probability distribution functions. The transmitted symbols, denoted by the row vector $\vect{x_i}$ of $\mat{x}$, are uniformly distributed, i.e., $f_X(\vect{x_i})=1/M$. 

The other channel we consider is the nonlinear fibre channel. The channel is modelled as an \ac{AWGN} channel whose noise power is dependent on the physical properties of the fibre transmission system, optical transmitted signal power and the \ac{ASE} noise power\cite{Splett1993Ultimate,Poggiolini2012GN}. Additionally, the modulation format used for transmission impacts the noise power, following \cite{Dar2013Properties,dar2014on}, and the \ac{SNR} of an optical fibre channel at optimum launch power can be predicted as
\begin{align}
    \text{SNR} &= \frac{P}{P_\text{ASE}+\eta_\text{tot}P^3} \\
    \eta_{\text{tot}}P^3 &\approx\left(\eta_1+\eta_2\Phi(X)\right)P^3\label{eq:Ch3_eta_tot} \\
    \text{SNR}_\text{opt} &\approx \frac{\left(\frac{\frac{1}{2\eta_1} P_\text{ASE}}{(1+\frac{\eta_2}{\eta_1}\Phi(X))}\right)^\frac{1}{3}}{P_\text{ASE}+\frac{P_\text{ASE}}{2}} \,,
\end{align}
 with the nonlinear coefficients $\eta_1$ and $\eta_2$, the launch power $P$, \ac{ASE} noise power $P_{\text{ASE}}$ and the excess kurtosis $\Phi(X) \triangleq \frac{\mathbb{E}\left[|X|^4\right]}{\mathbb{E}\left[|X|^2\right]^2}-2$ of the complex constellation. A closed form equation for $\eta_1$ can be found in \cite[Eq.~(5,10,11)]{Semrau2019AClosedForm} and $\eta_2$ can be found in \cite[Eq.~(16)]{Semrau2019AModulation}.

To simplify the expression for the performance of a optical transmission system given a transmitted constellation, the dependence on \ac{ASE} noise power cancels out when describing the expression as a ratio of the optimum \ac{SNR} of the input distribution and a reference distribution. Effectively lumping the channel description into two terms. Although any reference distribution can be chosen, the Gaussian distribution has zero excess kurtosis, i.e., $\Phi\left(X_\text{Gaussian}\right)=0$, and simplifies the expression even further. The ratio between the optimum SNR for the input distribution $X$ and the reference distribution $X_\text{ref}$ is then given as
\begin{align}
    \frac{\text{SNR}_\text{opt}}{\text{SNR}_\text{opt,ref}} &= \frac{\left(1+c\Phi(X)\right)^{-\frac{1}{3}}}{\left(1+c\Phi(X_\text{ref})\right)^{-\frac{1}{3}}} \\
    \text{SNR}_\text{opt} &= \left(1+c\Phi(X)\right)^{-\frac{1}{3}} \text{SNR}_\text{opt,Gaussian} \,, \label{eq:nonlinear_change}
\end{align}
with $\text{SNR}_\text{opt,Gaussian}=\frac{\left(\frac{1}{2\eta_1} P_\text{ASE}\right)^\frac{1}{3}}{\frac{3}{2}P_\text{ASE}}$ and eta-ratio $c = \frac{\eta_2}{\eta_1}$.
Eq.~\eqref{eq:nonlinear_change} is used to predict the change in \ac{SNR} when constellation $X$ is used compared to a reference constellation. 
This means, the channel performance as a function of the constellation is fully described with only $\text{SNR}_\text{opt,ref}$, $\Phi(X_\text{ref})$ and $c$. 
This expression will be used in Sec.~\ref{sec:nonlinear} to modify the signal power, effectively changing the \ac{SNR} as a function of the constellation.
Note that, if the modulation format changes, the launch power also changes to the optimum launch power for the new constellation and the other channel parameters remain  the same; Eq.~\eqref{eq:nonlinear_change} can be used to calculate the change in effective \ac{SNR}.
Therefore, the behaviour of the constellations designed for this channel is given by this compact expression, describing the channel with only the eta-ratio $c$ and the \ac{SNR} at optimum launch power when a reference distribution is transmitted.

\subsection{Achievable Information Rates}
 In this work, the amount of information that can be reliably transmitted is quantified using \acp{AIR}, specifically \ac{MI} or \ac{GMI}. The \ac{MI} \cite{Shannon1948Mathematical}, defined between $X$ and $Y$ is given as as \cite[Eq.~(8.47)]{Cover2005Elements}
\begin{equation}
    I(X;Y) = \mathop{{} \text{\Large $\mathbb{E}$}\!}_{X,Y} \left[ \log_2\left(\frac{f_{X,Y}(\vect{x},\vect{y})}{f_X(\vect{x})f_Y(\vect{y})} \right)\right] \,. \label{eq:mi_theory}
\end{equation}
For transmitting and receiving symbols, the \ac{MI} gives the maximum amount of information, measured in bits per symbol, that we can decode reliably.

In the case of a \ac{BICM} scheme, the GMI is a more appropriate measure as it predicts the performance of \ac{FEC} schemes\cite{Caire1988Bit-interleaved}. To calculate the \ac{GMI}, the sum of the mutual information between the input bits $C_k$ and received symbols $Y$ is used. Let $m=\log_2(M)$ be the number of bits transmitted per symbol, a vector of $m$ bits $\vect{c}_i =[c_{i,1},\ldots,c_{i,m}]^T$ is mapped to a corresponding symbol $\vect{x}_i$. 
The \ac{GMI} is given as\cite{Alvarado2018Achievable}

\begin{align}
    G &= \sum_{k=1}^m I(C_k;Y) \nonumber \\
    &= \sum_{k=1}^m \mathop{{} \text{\Large $\mathbb{E}$}\!}_{C_k,Y} \left[\log_2\frac{f_{Y|C_k}(\vect{y}|c_k)}{f_{Y}(\vect{y})} \right] \,. \label{eq:gmi_theory}
\end{align}

An \ac{AIR} when using a given \ac{FEC} scheme as inner code can be computed as
\begin{equation}
    R^* = m R (1-H_\text{b}\big(\text{BER})\big) \label{eq:R_star}\,,
\end{equation}
where $m$ is the number of transmitted bits per 2D, $R$ the inner \ac{FEC} code rate and $\big(1-H_\text{b}(p)\big) = 1 + p \log_2 p + (1 - p) \log_2 (1 - p)$ is the capacity of the binary symmetric channel with cross-over probability $p$ determined by the binary input of the FEC encoder and the binary output of the FEC decoder. The decoder output is here assumed to contain independent errors after the use of an ideal interleaver. The BER is obtained with error counting after the soft-decision \ac{FEC} decoder. 
In the remainder of this paper, this rate is used as a performance metric, together with \ac{MI} and \ac{GMI}.

We use \ac{MI} and \ac{GMI} to refer to the quantities defined in Eq.~\eqref{eq:mi_theory} and Eq.~\eqref{eq:gmi_theory}, respectively.  The generic \ac{AIR} label is used whenever different information rates are shown (including  MI, GMI and $R^*$).

\subsection{Nonlinear Optimisation Methods Based on Gradients}
\label{sec:gradient_descent}

For the optimisation problem, i.e., finding the locations of constellation points that maximise throughput, we used a gradient descent algorithm.
Optimising the constellations for \ac{GMI} is a nonconvex problem and therefore the optimsation is affected by converging to a non-global minima. Although for nonconvex problems the derivative driven algorithms are not guaranteed to converge to global optima, we have have obtained good results. We have compared both ac{BFGS} \cite{Broyden1970Convergence,Fletcher1970Approach,Goldfarb1970Family,Shanno1970Conditioning} and trust-region optimisation algorithm and observed fewer local minima from the trust-region. 
In this work,we focused on the trust-region algorithms\cite{Coleman1996,Byrd1987Trust,Byrd1996Analysis} where the Jacobian of the Jacobian, known as the Hessian, is estimated using a symmetric rank-one method\cite{Conn1991Convergence} as described in \cite[p.~146]{Nocedal2006Numerical}. The optimisation step is described by 
\begin{align}
    \vect{x}_{k+1} &= \vect{x}_k + \vect{s}_k \,,
\end{align}
where $\vect{x}_k$ is the constellation at step $k$ and the update $\vect{s}_k$ is determined by the trust-region sub-problem defined as
\begin{align}
    \min_{\vect{s}_k}&\, f_k + \vect{g}_k^T \vect{s}_k + \frac{1}{2}\vect{s}_k^T \mat{B}_k \vect{s}_k \nonumber\\
    &\text{s.t. }||\vect{s}_k|| \leq \Delta_k \label{eq:trustregion} 
    \,,
\end{align}
where $f_k$ is the function value, $\vect{g}_k$ is the gradient and $\mat{B}_k$ the inverse of the Hessian at $\vect{x}_k$. The trust region $\Delta_k$ limits the size of the step $\vect{s}_k$ which shrinks as we get close to optimum by the algorithm. We initialise this parameter of the algorithm as 1. For this method, we need an accurate gradient as it is used for both the step and the Hessian estimation. The next step is to solve the step sub-problem Eq.~\eqref{eq:trustregion}, for which we use Steihaug's conjugate gradient method\cite{Steihaug1983Conjugate}. The sub-problem is a simpler problem and easier to optimise than the entire optimisation problem and the sub-problem is solved at every step. Finding a solution for the sub-problem is often less computationally expensive than a single gradient calculation of the \ac{AIR} with respect to the input constellation.
A fast trust-region optimisation algorithm is possible via the improvements proposed in Sec.~\ref{sec:derivative}, where we describe the  efficient gradient calculation proposed in this work.

\section{Efficient Gradient Computation of MI and GMI for the AWGN Channel}
\label{sec:derivative}

One of the difficulties of calculating the gradients of \ac{MI} and \ac{GMI} is that, for a fixed \ac{AWGN} variance, the input constellation needs to be constrained to constant average power. The most straightforward approach to constellation optimisation is to use the finite-difference method, renormalising the constellation after every finite-difference exploration step for the gradient calculation.
However, the renormalisation has the consequence that, without extra manipulation, the entire objective function needs to be recalculated, for every constellation point, for every dimension. 
This is because, if the location of a single constellation point is changed, all the points need to be rescaled to maintain the same average power.

In this work, we propose an efficient method of obtaining the gradient by splitting the problem into two functions in a composition. The first function normalises the constellation --- that is, it takes the constellation points as an input and outputs a vector of normalised constellation points.  The output of the first function is then passed to a function that is either an unconstrained \ac{MI} or \ac{GMI} function. This maps a vector with all constellation points to a single \ac{AIR} value, in this work we use Eq.~\eqref{eq:mi_theory} and Eq.~\eqref{eq:gmi_theory}.

Additionally, after some algebraic transformation of the complete objective function, the gradient can be expressed as a rearrangement of the terms used to calculate the \ac{MI} and \ac{GMI}. The integral calculating the expectation of the Gaussian-noise variable still needs to be evaluated, which can be done efficiently using a \ac{GHQ} \cite{Alvarado2018Achievable}.

\subsection{Mutual Information (MI)}
\label{sec:mi_grad}
We start by analysing the MI calculation for the \ac{AWGN} channel. The MI and GMI calculations as presented in this work are similar and the derivation for the simpler MI calculation can be taken over to the GMI derivation. Let the MI function be \cite[Eq.~(21)]{Alvarado2018Achievable}
\begin{align}
    I &= m- \frac{1}{M}\sum_{i=1}^M  \E g_i(\vect{z},\mat{x}_\mathcal{J}) \dz \,, \label{eq:MIcalc}
\end{align} 
where
\begin{align}
    f(\vect{z}) &= \frac{1}{\left(\pi\sigma_z^2\right)^N}\exp\left(\frac{||\vect{z}||^2}{-\sigma_z^2}\right) \\
    g_i(\vect{z},\mat{x}_\mathcal{J}) &\triangleq \log_2\left( \sum_{j\in\mathcal{J}} h(\vect{z},\vect{x}_i,\vect{x}_j) \right) \\
    h(\vect{z},\vect{x}_i,\vect{x}_j) &\triangleq \exp \left(\frac{||\vect{x}_i-\vect{x}_j||^2 + 2 \langle \vect{z}, (\vect{x}_i-\vect{x}_j)\rangle}{-\sigma_z^2}\right) \label{eq:hfunc} \,,
\end{align}
with $\mathcal{J}=\{1,\ldots,M\}$, $\vect{z}=\vect{y}-\vect{x}_i$, $\vect{d}_{ij}=-\vect{d}_{ji}=\vect{x}_i-\vect{x}_j$ and $\sigma_z^2$ the noise variance per 2 real dimensions. The variable $Z$ is a substitution to split the additive noise from the transmitted symbol for the expected value calculation.

Following the steps in Appendix~\ref{ap:mi_derivation}, we obtain:
\begin{equation}
    \nabla I = [\ddx{\vect{x}_1}I, \ddx{\vect{x}_2}I,\ldots,\ddx{\vect{x}_M}I ], \label{eq:MIgrad} \\
\end{equation}%
where
\begin{multline}
\ddx{\vect{x}_n}I = -\frac{1}{M} \E \Bigg[ \frac{\sum_{j\in\setJn} \left(\frac{2\vect{d}_{nj} + 2 \vect{z}}{-\sigma_z^2}\right) h(\vect{z},\vect{x}_n,\vect{x}_j) }{\log(2)\sum_{j\in\mathcal{J}} h(\vect{z},\vect{x}_n,\vect{x}_j) }\\
+\sum_{i\in\setJn}\frac{-\left(\frac{2\vect{d}_{in} + 2 \vect{z}}{-\sigma_z^2}\right) h(\vect{z},\vect{x}_i,\vect{x}_n) }{\log(2)\sum_{j\in\mathcal{J}} h(\vect{z},\vect{x}_i,\vect{x}_j) } \Bigg] \dz \label{eq:ddxnMI} \,,
\end{multline}
and $\setJn= \mathcal{J}\setminus\{n\}=\{j\in \mathcal{J}; j\neq n\}$.

We can simplify the computation of Eq.~\eqref{eq:ddxnMI} by using the symmetry in $\vect{z}$ and $\vect{d}_{ij}$
\begin{align}
    \ddx{\vect{x}_n}g_i(\vect{z},\mat{x}_\mathcal{J}) =  \ddx{\vect{x}_i}g_n(\vect{z},\mat{x}_\mathcal{J}) \,.
\end{align}
To calculate the gradient using Eq.~\eqref{eq:ddxnMI}, we only need $M^2$ of the subfunction Eq.~\eqref{eq:hfunc}, all of which is already calculated for the \ac{MI} calculation. For the finite-differences approach, at least $2N\times M$ more are needed.

Then, to obtain the gradient we need to calculate the multidimensional integral over the support of $\vect{z}$ in Eq.~\eqref{eq:ddxnMI}. To compute the \ac{MI} and its gradient in the \ac{AWGN} channel, we can use the \ac{GHQ} to efficiently numerically evaluate the integral \cite[Sec.~IV.B]{Alvarado2018Achievable}. 

\subsection{Generalised Mutual Information (GMI)}
For the bit-wise decoder commonly used in \ac{BICM} channels, the \ac{GMI} is an accurate predictor of the post-FEC performance. To optimise the \ac{GMI} \cite[Eq.~(22)]{Alvarado2018Achievable}, we compute its gradient similarly to what done for Eq.~\eqref{eq:ddxnMI}. We then start from expanding Eq.~\eqref{eq:gmi_theory} to
\begin{multline}G =  m- \frac{1}{M} \sum_{b\in\{0,1\}} \sum_{i\in\setIbk} \sum_{k=1}^m \E \\
   \log_2\left(\frac{\sum_{j\in\mathcal{J}} h(\vect{z},\vect{x}_i,\vect{x}_j)}{\sum_{p\in\setIbk} h(\vect{z},\vect{x}_i,\vect{x}_p)}\right)
    \dz \label{eq:gmi_function} \,,
\end{multline} 
where $\setIbk = \{i\in\mathcal{J}; c_{i,k}=b \}$ is the set of indices where the $k$-th bit is $b$.

Eq.~\eqref{eq:gmi_function} can be rewritten to be more similar to Eq.~\eqref{eq:MIcalc} and derive a gradient similarly to Eq.~\eqref{eq:ddxnMI}:
\begin{multline}
   G = m- \frac{1}{M} \sum_{b\in\{0,1\}} \sum_{i\in\setIbk} \sum_{k=1}^m \E \\
   \left[ \log_2\left(\sum_{j\in\mathcal{J}} h(\vect{z},\vect{x}_i,\vect{x}_j)\right) \right.\\
   \left.-\log_2\left(\sum_{p\in\setIbk} h(\vect{z},\vect{x}_i,\vect{x}_p)\right) \right] 
    \dz \,. \label{eq:GMIcalc}
\end{multline}

Now after introducing a new subfunction where the sum is dependent on the bit mapping, $\setIbk$,:
\begin{align}
    g_i(\vect{z},\mat{x}_{\setIbk}) &\triangleq \log_2\left( \sum_{p\in\setIbk} h(\vect{z},\vect{x}_i,\vect{x}_p) \right) \,,
\end{align}
for which the partial derivative can be calculated equivalently to Appendix~\ref{ap:mi_derivation}, Eq.~\eqref{eq:partial_g}, we can formulate the gradient for the \ac{GMI} as
\begin{align}
\begin{aligned}
    \ddx{x_n}G = -\frac{-m}{M} &\E \ddx{\vect{x}_n}  g_i(\vect{z},\mat{x}_{\mathcal{J}})\dz \\
    - \frac{m}{M}\sum_{i \in \setJn}  &\E\ddx{\vect{x}_{n\in J}} g_i(\vect{z},\mat{x}_{\mathcal{J}})\dz \\
    - \frac{1}{M} \sum_{k=1}^m     &\E \ddx{\vect{x}_n}  g_i(\vect{z},\mat{x}_{\setIbk})\dz \\
    - \frac{1}{M} \sum_{k=1}^m     &\sum_{i\in \settIbk} \E \ddx{\vect{x}_{n\in J}}  g_i(\vect{z},\mat{x}_{\setIbk})\dz
\end{aligned}\label{eq:gmi_derrivative} \,,
\end{align}
where $\settIbk = \{i\in\setIbk;i\neq n \}$.

Following the steps from Sec.~\ref{sec:mi_grad}, the gradient of the \ac{GMI} is:
\begin{equation}
    \nabla G = [\ddx{\vect{x}_1}G, \ddx{\vect{x}_2}G,\ldots,\ddx{\vect{x}_M}G ] \\
\end{equation}%
\begin{multline}
\ddx{\vect{x}_n}G = -\frac{1}{M} \E \Bigg[\\
m\frac{\sum_{j\in\setJn} \left(\frac{2\vect{d}_{nj} + 2 \vect{z}}{-\sigma_z^2}\right) h(\vect{z},\vect{x}_n,\vect{x}_j) }{\log(2)\sum_{j\in\mathcal{J}} h(\vect{z},\vect{x}_n,\vect{x}_j) }\\
+m\sum_{i\in\setJn}\frac{-\left(\frac{2\vect{d}_{in} + 2 \vect{z}}{-\sigma_z^2}\right) h(\vect{z},\vect{x}_i,\vect{x}_n) }{\log(2)\sum_{j\in\mathcal{J}} h(\vect{z},\vect{x}_i,\vect{x}_j) } \\
 + \sum_{k=1}^m \frac{\sum_{j\in\settIbk} \left(\frac{2\vect{d}_{nj} + 2 \vect{z}}{-\sigma_z^2}\right) h(\vect{z},\vect{x}_n,\vect{x}_j) }{\log(2)\sum_{j\in\mathcal{J}} h(\vect{z},\vect{x}_n,\vect{x}_j) }\\
+ \sum_{k=1}^m \sum_{i\in\settIbk}\frac{-\left(\frac{2\vect{d}_{in} + 2 \vect{z}}{-\sigma_z^2}\right) h(\vect{z},\vect{x}_i,\vect{x}_n) }{\log(2)\sum_{j\in\mathcal{J}} h(\vect{z},\vect{x}_i,\vect{x}_j) }\Bigg] \dz \,. \label{eq:gmi_gradient}
\end{multline}

Now, again, the only step left to obtain the gradient is to evaluate the expected value of the multidimensional integral over the support of $\vect{z}$, which again can be numerically evaluated with a \ac{GHQ}.

\section{Channel Specific Constellation Optimisation}
\label{sec:chain_rule}

In the previous section, we simply computed the gradient of the MI and GMI for the \ac{AWGN} channel. This allows us to compute the performance metric given the constellation coordinates and fixed noise variance. For optimisation of constellations in the \ac{AWGN} channel, the \ac{SNR} is fixed at the desired value. When looking at the definition of $\text{SNR} \triangleq \frac{\mathrm{E}[||X||^2]}{\mathrm{E}[||Z||^2]}$, and assuming that the $\mathrm{E}[||Z||^2]\triangleq N\sigma_z^2$ is fixed, it can be seen that the signal power $\mathrm{E}[||X||^2]$ needs to be constrained, as previously noted.
Normalising the constellation leads to an unconstrained optimisation problem which is invariant to the input constellation energy. This invariance is achieved by introducing a normalisation function and defining the optimisation function as the composite of the normalisation and the \ac{AIR}, as described in the next section.

\subsection{Gradient chain rule for the AWGN Channel}

For the AWGN channel, the only extra function we need to introduce is the normalisation function.
We can define the normalisation function as
\begin{equation}
        \mat{u}(\mat{x}) = \frac{\mat{x}}{\sqrt{\frac{1}{MN}} ||\mat{x}||_F} \,,  \label{eq:normalise}
\end{equation}
whose Jacobian is given by
\begin{align}
    \mat{J}_{\vect{u}} &=  \frac{||\mat{x}||_F^2 \mathbf{I}_{M,2N}-\mat{x} \otimes \mat{x}}{\sqrt{\frac{1}{MN}}||\mat{x}||_F^3 }\\
    (\mathbf{I}_{M,2N})_{i,j,k,l} &= \left\{\begin{aligned}
        &1 & \quad &\text{if } i=j \wedge k=l\\
        &0 & \quad &\text{otherwise}
    \end{aligned}\right.\\
    (\mat{x}\otimes\mat{x})_{i,j,k,l} &= \vect{x}_{i,j}\vect{x}_{k,l} \,,
\end{align}
for the complex $M \times 2N$ vector $\vect{x}\in\mathbb{R}^{M\times 2N}$ and $\mat{J},\mat{I}\in \mathbb{R}^{M,D,M,2N}$.
With this function and its Jacobian, we can use the following objective function and apply the chain rule to obtain the gradient
\begin{align}
    o(\vect{x}) &= f\big(\mat{u}(\vect{x})\big) \label{eq:chain_optimisation}\\
    \nabla  o(\vect{x}) &= \left(\mat{J}_o(\vect{x})\right)^H = \Big(\mat{J}_f\big(\mat{u}(\vect{x})\big)\mat{J}_{\mat{u}}(\vect{x})\Big)^H \label{eq:chain_jacobian} \,,
\end{align}
which can be used in the optimisation method. Here, we can use the \ac{MI} from Eq.~\eqref{eq:MIcalc} as $f(\vect{x})$ and its gradient, i.e., Eq.~\eqref{eq:MIgrad}, as $\mat{J}_f(\vect{x})$.

In summary, for an \ac{MI} calculation, first the constellation is normalised using Eq.~\eqref{eq:normalise}. Next, we calculate the contribution to the GMI and the gradient for each constellation point separately. Starting from $x_i$ we pre-calculate
\begin{equation}
    h(\vect{z},\vect{x}_i,\vect{x}_j) = \exp \left(\frac{||\vect{d}_{ij}||^2 + 2 \langle \vect{z} , \vect{d}_{ij}\rangle}{-\sigma_z^2}\right)
\end{equation}
for all $x_j$ and sample points $z$, n.b., if $x_i = x_j$, the result is $1$. The MI is calculated by first summing tributaries $h(\vect{z},\vect{x}_i,\vect{x}_j)$ over $j$ and then, after taking the log, taking the (weighted) sum over $z$. For the gradient calculation, start by summing all tributaries of $h(\vect{z},\vect{x}_i,\vect{x}_j)$ over $j$, then multiplying by $2(\vect{z}+\vect{d_{ij}})$ and finally take the (weighted) sum over $z$. Sum the contributions over all $i$ and scale with constants according to the integration method chosen.  

\subsection{Gradient chain rule for the Nonlinear Fibre Channel}
\label{sec:nonlinear}

To extend the method to optimise the constellation for the nonlinear fibre channel, the normalisation function is extended to reflect the change in SNR as a function of the transmitted constellation described in section~\ref{sec:channel_model}. 
When calculating the \ac{AIR} for the nonlinear fibre channel, the performance of the constellation is evaluated at optimum launch power. Assuming a known \ac{SNR} for a reference constellation at know launch power, the constellation can be scaled to reflect the \ac{SNR} change due to the fibre nonlinearity compared to the reference constellation. For convenience, a Gaussian distribution is used as the reference constellation, as shown in Eq.~\eqref{eq:nonlinear_change}. The \ac{AIR} can then be calculated using either Eq.~\eqref{eq:MIcalc} or Eq.~\eqref{eq:GMIcalc} with the scaled constellation as the input.
This is performed analogously to the normalisation function Eq.~\eqref{eq:normalise} and its effect on the gradient Eq.~\eqref{eq:chain_optimisation} is described as:
\begin{align}
     o(\mat{x}) &= f(\mat{n}(\mat{u}(\mat{x}))) \label{eq:nonlin_objective} \\
     \mat{n}(\mat{x}) &= \mat{x}\sqrt{\left(1+c\Phi\left(\mat{x}\right)\right)^{-\frac{1}{3}}} \,,
\end{align}
reflecting the amplitude change as described in Eq.~\eqref{eq:nonlinear_change}.

We can now define functions $v(\mat{x})$ and $w(\mat{x})$ to calculate its Jacobian of $\mat{n}(\mat{x})$:
\begin{align}
    \mat{n}(\mat{x}) &= \mat{x}\left(1+c\left(\frac{v(\mat{x})}{w(\mat{x})}-2\right)\right)^{-\frac{1}{6}} \\
    v(\mat{x}) &= \frac{1}{M}\sum_{j=1}^M \big(\Re\{x_j\}^2 +\Im\{x_j\}^2\big)^2 \\
    w(\mat{x}) &= \left( \frac{1}{M}\sum_{j=1}^M \big(\Re\{x_j\}^2 +\Im\{x_j\}^2\big) \right)^2 
\end{align}
for which we can find derivatives
\begin{align}
    v^\prime(\mat{x}) &= \ddx{x_i} v(\mat{x}) = \frac{4x_i}{M} \Big(\Re\{x_j\}^2 +\Im\{x_j\}^2\Big) \\
    w^\prime(\mat{x}) &= \ddx{x_i} w(\mat{x}) =  \frac{4x_i}{M} \left( \frac{1}{M}\sum_{j=1}^M \big(\Re\{x_j\}^2 +\Im\{x_j\}^2\big) \right) \,.
\end{align}
Then, we can obtain the Jacobian in two steps, first the diagonal
\begin{multline}
    \ddx{x_i} n_i(\mat{x}) = \left(1+c\left(\frac{v(\mat{x})}{w(\mat{x})}-2\right)\right)^{-\frac{1}{6}} \\
    -\frac{x_i}{6}\left(1+c\left(\frac{v(\mat{x})}{w(\mat{x})}-2\right)\right)^{-\frac{7}{6}}\\ \cdot\left(\frac{v^\prime(\mat{x}) w(\mat{x})- v(\mat{x}) w^\prime(\mat{x})}{w(\mat{x})^2}\right) \,.
\end{multline}
Now for all elements not on the diagonal, i.e. $i \neq p$, we obtain:
\begin{multline}
    \ddx{x_i} n_p(\mat{x}) =  -\frac{x_p}{6}\left(1+c\left(\frac{v(\mat{x})}{w(\mat{x})}-2\right)\right)^{-\frac{7}{6}} \\
    \cdot\left(\frac{v^\prime(\mat{x}) w(\mat{x})- v(\mat{x}) w^\prime(\mat{x})}{w(\mat{x})^2}\right) \,,
\end{multline}
with $|x_j|^2 = \Re\{x_j\}^2 +\Im\{x_j\}^2$ and
\begin{multline}
    \frac{v^\prime(\mat{x}) w(\mat{x})- v(\mat{x}) w^\prime(\mat{x})}{w(\mat{x})^2}\\ = \frac{4M x_i\left(|x_i|^2\sum_{j=1}^M |x_j|^2 - \sum_{j=1}^M |x_j|^4  \right)}{\left(\sum_{j=1}^M |x_j|^2\right)^3} \,.
\end{multline}
Note that because of the dimensionality of $\mat{x}\in\mathbb{C}^{M,1}$ each individual symbol $x_i$ is here a complex scalar. To scale the current model beyond a single dimension needs further work \cite{Liga2020Extending}.

This result can be used to extend Eq.~\eqref{eq:chain_optimisation} to include the expected change in \ac{SNR} from reduced nonlinear distortion due to the changed excess kurtosis of the constellation. This new objective function then finds constellations where the trade-off between shaping and nonlinearity is optimised.

\section{Results for the AWGN Channel}\label{sec:results}

In this section, the performance of the optimised constellations in the \ac{AWGN} channel is shown. The optimisation process is described in detail in appendix~\ref{ap:algorithm}. In Fig.~\ref{fig:opt_traj}, an example of a constellation undergoing the optimisation process is shown. The optimisation trajectory for 12~dB \ac{SNR} using a regular 64-\ac{QAM} as a starting constellation is shown. The 64-\ac{QAM} constellation is chosen for clarity of illustration of the optimisation trajectory, due to its relatively low cardinality. The constellation points generally do not have a linear optimisation trajectory, but the optimisation does result in a regular-looking constellation. For the example constellation, the result resembles an \ac{APSK} constellation where the inner amplitude ring has two constellation points for every value of phase shift.

\begin{figure}[thb]
    \centering
\begin{tikzpicture}

\definecolor{startcolor}{RGB}{120,217,112}
\definecolor{endcolor}{RGB}{63,37,0}

\begin{axis}[%
width=0.75\columnwidth,
scale only axis,
point meta min=3.7768,
point meta max=3.9491,
xmajorgrids,
ymajorgrids,
axis equal image=true,
colormap={mymap}{rgb255=(120,217,112) rgb255=(63,205,153) rgb255=(56,166,206) rgb255=(94,110,226) rgb255=(142,59,186) rgb255=(158,33,105) rgb255=(125,31,25) rgb255=(63,37,0)},
colorbar,
colorbar style={
    ylabel={GMI [bit/2Dsym]},
    ytick distance=0.02,
    height=4.5cm,
    at={(1.05,0)},
    anchor=south west,
    },
legend pos=outer north east,
legend cell align=left,
legend style={font=\footnotesize},
table/search path={data/optimisation},
]

 \addplot [color=startcolor, thick, only marks, mark=*, mark options={solid, startcolor}]
   table[]{opt_start.tsv};
   \addlegendentry{Starting point}
   
   \addplot[mesh,thick,forget plot,empty line=jump]
table[point meta=\thisrow{c}] {%
opt_traj.tsv};

\addplot [color=endcolor,thick, only marks, mark=o, mark options={solid, endcolor}]
  table[]{opt_end.tsv};
  \addlegendentry{Optimised}
\end{axis}
\end{tikzpicture}
    \label{fig:opt_traj}
\end{figure}

Next, we describe at the results for 2D and 4D geometrically shaped constellations for the \ac{AWGN} channel. To obtain these results, we followed the trust-region method of Sec.~\ref{sec:gradient_descent} with the GMI gradient in Eq.~\eqref{eq:gmi_gradient} and the normalisation for the AWGN channel Eq.~\eqref{eq:normalise}.
The optimisation is carried out by starting with different initial constellations. Two fixed constellations; namely, a regular \ac{QAM} constellation and one designed following \cite{Meric2015Approaching}, as well as randomly generated starting points. In the optimisation, symmetry around every axis was imposed, akin to the \ac{BRGC}, so that the most significant bit describes which half of the plane the constellation point is in and the other half of the points are a mirrored copy. As both the regular \ac{QAM} and the constellations from  M\'eric~\cite{Meric2015Approaching} already meet these requirements, these constellations can be directly used as a starting point for these optimisations. We have found the solutions with these constraints performed equally well as solutions that allowed asymmetries in the region where the MI/GMI are close to the maximum entropy of the constellations. By using Eq.~\eqref{eq:chain_optimisation}, we can obtain more accurate gradients with fewer iterations. This allows the optimisation of high cardinality constellations, which in this paper were selected to be up to 8192 points.

\subsection{2D Constellations Shaping for the AWGN Channel}

The performance of the \ac{GS-2D} constellations tailored to the \ac{AWGN} channel is shown in  Fig.~\ref{fig:optimised_2D_from_meric} and is illustrated in terms of gap to \ac{AWGN} capacity at the \ac{SNR} the constellation is optimised for. 
The optimisation was performed for $m=3,4,\ldots,13$, and for integer values of \ac{SNR} in the range shown. It can be seen that the performance of higher-cardinality constellations in terms of gap to capacity improves compared to the results obtained for lower-order modulation formats.

The capacity of the \ac{AWGN} channel is $\log_2(1 + \mathrm{SNR})$ bit/2D  \cite{Shannon1948Mathematical}. One would expect the \ac{MI} and \ac{GMI} to grow in line with the capacity until the throughput is limited by the cardinality of the constellation. Therefore, before saturation is reached the gap between \ac{GMI} and the \ac{AWGN} capacity falls within a relatively small range compared to the achieved \ac{GMI}. Showing the performance of the optimised constellations as the gap to the \ac{AWGN} capacity improves the presentation of the results, plotted over a wide range of \acp{SNR}.

\begin{figure}[thb]
    \centering
    \begin{tikzpicture}
         \begin{axis}[
        snrgap,
        colormap name=isoL-20,
        cycle list={[ of colormap]},
        width=\linewidth,
        ymin=0,ymax=0.2,
        enlargelimits=false,
        y tick label style={/pgf/number format/.cd,%
          scaled y ticks = false,
          fixed},
     ]
     \foreach \i in {8,16,32,64,128,256,512,1024,2048,4096,8192} {
     \addplot+[thick] table[x=SNR,y expr=\thisrow{C}-\thisrow{QAM\i}] {data/awgn_opt.txt};
     }

     \node[qamlabel] at (axis cs:4.6,0.175) {8};
     \node[qamlabel] at (axis cs:8.5,0.175) {16};
     \node[qamlabel] at (axis cs:11,0.175) {32};
     \node[qamlabel] at (axis cs:14,0.175) {64};
     \node[qamlabel] at (axis cs:17,0.175) {128};
     \node[qamlabel] at (axis cs:20,0.175) {256};
     \node[qamlabel] at (axis cs:22.25,0.175) {512};
     \node[qamlabel] at (axis cs:26,0.175) {1024};
     \node[qamlabel] at (axis cs:28,0.1625) {2048};
     \node[qamlabel] at (axis cs:28,0.105) {4096};
     \node[qamlabel] at (axis cs:28,0.065) {8192};
     
     \end{axis}
    \end{tikzpicture}     \caption{Gap to capacity in GMI for optimised GS-2D constellations tailored to the AWGN channel enforcing 2 axes of symmetry for every round number SNR\@. Numbers on the lines indicate the constellation order.}
    \label{fig:optimised_2D_from_meric}
\end{figure}
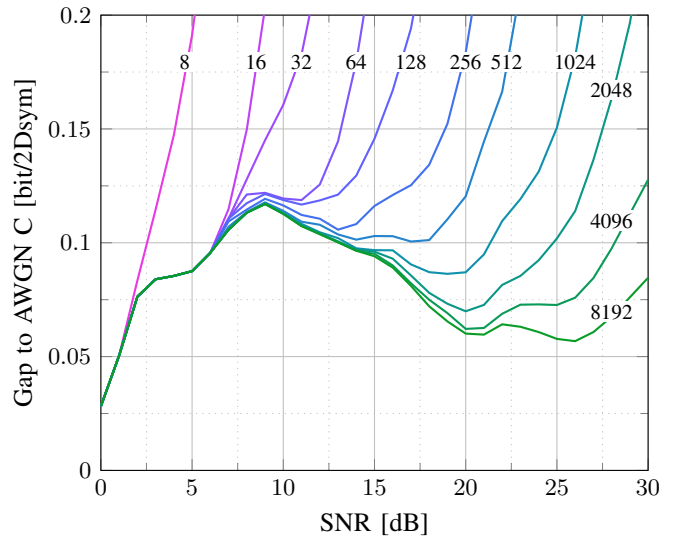

\begin{figure}[thb]
    \begin{tikzpicture}
    \begin{axis}[
        width=\linewidth,
        xmin=0,xmax=30,
        ymin=0,ymax=0.6,
        cycle list={%
            Set1-A,
            Set1-B,
            Set1-C,
            Set1-D,
            Set1-I,
        },
        snrgap,
        legend pos=north west,
        ]
        \addplot+[very thick] table[x=SNR, y expr=\thisrow{C}-\thisrow{M8192}] {data/opt2Dgmi.txt};
        \addplot table[x=SNR,y=square] {data/mingapGMI.txt};
        \addplot table[x=SNR,y=suntilborg] {data/mingapGMI.txt};
        \addplot table[x=SNR,y=meric] {data/mingapGMI.txt};
        \addplot table[x=SNR, y=gap] {data/uniMI.txt};
        \legend{GS-2D-8192, Envelope Square QAM, Envelope Sun-Tilborg, Envelope M\'eric, MI uniform $\infty$-QAM}
    \end{axis}
    \end{tikzpicture}     \caption{Gap to capacity in GMI for optimised quadrant-symmetric 8192-ary constellation compared to the envelopes of regular square QAM, Sun-Tilborg~\cite{Sun1993Approaching} and M\'eric~\cite{Meric2015Approaching}. The MI for uniform $\infty$-QAM is shown a reference for the 1.53~dB shaping gain\cite{Forney1984Effecient}.
    }\label{fig:result_highlight}
\end{figure}
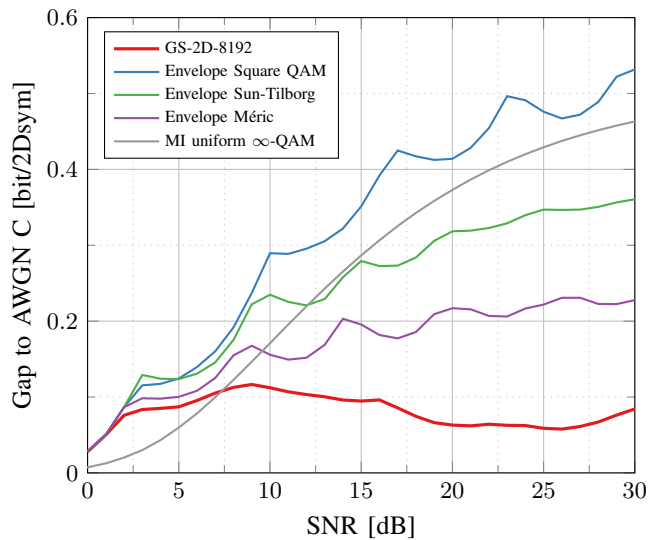

In Fig.~\ref{fig:result_highlight}, the performance of the 2-dimensional optimised constellation is shown for the highest cardinality constellation $M=8192$ investigated in this work. This performance is compared to the one of benchmark constellations such as square \ac{QAM}, Sun-Tilborg~\cite{Sun1993Approaching} and M\'eric constellations~\cite{Meric2015Approaching}, the latter two being capacity-approaching in terms of \ac{MI}. The cardinality of the benchmark constellations was varied based on the selected \ac{SNR} (up to $M=8192$) and the performance of the resulting envelopes is shown in Fig.~\ref{fig:result_highlight}. 
The MI for uniform $\infty$-QAM is shown as a reference and it can be seen that it saturates to the 1.53~dB (0.51 bit) gap \cite{Forney1984Effecient}. The optimised $M=8192$ gives the best  performance of all cardinalities considered. For the optimisation method presented in this paper, the higher cardinality constellation will always achieve the performance of a lower cardinality constellation by taking the lower cardinality constellation and doubling each constellation point (two constellation points having the same coordinates), effectively ignoring the transmitted bit which selects between two constellation points with identical coordinates. 
It can be seen that the optimised constellations from this work outperform all other constellations in terms of GMI for all SNR values between 0 and 30 dB. The GMI of our optimal constellation also outperforms the MI of uniform $\infty$-QAM for SNR \textgreater 7 dB.

\subsection{4D Constellations Shaping for the AWGN Channel}
\label{sec:shape4d}

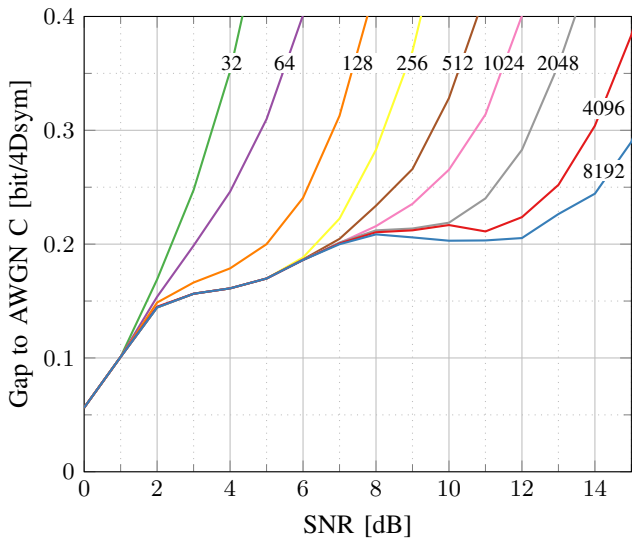
\begin{figure}[thb]
    \centering
    \begin{tikzpicture}
         \begin{axis}[
        snrgap,
        ylabel={Gap to AWGN C [bit/4Dsym]},
        cycle list={%
            Set1-A,
            Set1-B,
            Set1-C,
            Set1-D,
            Set1-E,
            Set1-F,
            Set1-G,
            Set1-H,
            Set1-I
        },
        width=\linewidth,
        ymin=0,ymax=0.4,
        xmin=0,xmax=15,
        enlargelimits=false,
        y tick label style={/pgf/number format/.cd,%
          scaled y ticks = false,
          fixed},
     ]
        \pgfplotsset{cycle list shift=2}
     \foreach \i in {32,64,128,256,512,1024,2048,4096,8192} {
     \addplot+[thick] table[x=SNR,y expr=2*\thisrow{C}-\thisrow{QAM\i}] {data/awgn_opt_4d.txt};
     }
     \node[qamlabel] at (axis cs:4.05,0.35) {32};
     \node[qamlabel] at (axis cs:5.5,0.35) {64};
     \node[qamlabel] at (axis cs:7.5,0.35) {128};
     \node[qamlabel] at (axis cs:9,0.35) {256};
     \node[qamlabel] at (axis cs:10.25,0.35) {512};
     \node[qamlabel] at (axis cs:11.5,0.35) {1024};
     \node[qamlabel] at (axis cs:13,0.35) {2048};
     \node[qamlabel] at (axis cs:14.25,0.31) {4096};
     \node[qamlabel] at (axis cs:14.25,0.255) {8192};

     \end{axis}
    \end{tikzpicture}     \caption{GMI-capacity gap for optimised GS-4D constellations tailored to the AWGN channel enforcing 4 axes of symmetry for every round number SNR\@. Numbers on the lines indicate the constellation order $M$.}
    \label{fig:optimised_4D_from_meric}
\end{figure}

The significant improvements in computation efficiency now allow us to extend the optimisation of constellations for the \ac{AWGN} channel to 4D, something which has been challenging to achieve to date. In Fig.~\ref{fig:optimised_4D_from_meric}, the results of optimising constellations for the 4D \ac{AWGN} channel are shown. 
The results follow a similar trend to the results presented for 2D, however, we can see that a cardinality of 8192 is not enough to decrease the gap to capacity for higher SNR values. 

Next, we compare the performance of the designed constellations with the recently published 4D constellation with symmetry constraints, presented in \cite{chen2020analysis}, which the authors named \ac{OS}. The authors showed how it was possible to reduce the 128-ary constellation to just 5 variables. The 128$\times$4 variables describing all points in all dimensions were reduced to 32 by enforcing symmetry and that subspace was then intuitively spanned using 5 variables. In contrast, the method introduced in this paper greatly improves the efficiency of the constellation design and allows the optimisation of all variables within the orthant. In the case of 128 constellation points this is 32 points with 4 dimensions each, totalling 128 variables. 

For this comparison we used the same simulation parameters from \cite{chen2020analysis}; \ac{SSFM} simulation\cite{Menyuk1987Nonlinear} of 80-km fibre spans with 100 logarithmically distributed steps, 0.21~dB/km attenuation, chromatic dispersion of 16.9~ps/(nm$\cdot$km) with a nonlinear coefficient $\gamma$ of 1.32~/(W$\cdot$km), amplified with a 4.5~dB noise figure optical amplifier. The transmitted signal consisted of 11 channels of 45~Gbd, spaced at 50~GHz. All transmitted channels had different randomly generated data and only the centre channel was demodulated. A root-raised cosine filter with 1\% roll-off was used for matched filtering. The fibre propagation was emulated by using the \ac{SSFM} with a fixed step size of 100~m. The post-\ac{FEC} results were obtained using the $R=4/5$ DVB-S2 \ac{LDPC} decoded with 200 iterations.
For each point, the optical launch power was swept in steps of 0.5~dB to find the maximum \ac{SNR} as a function of distance.

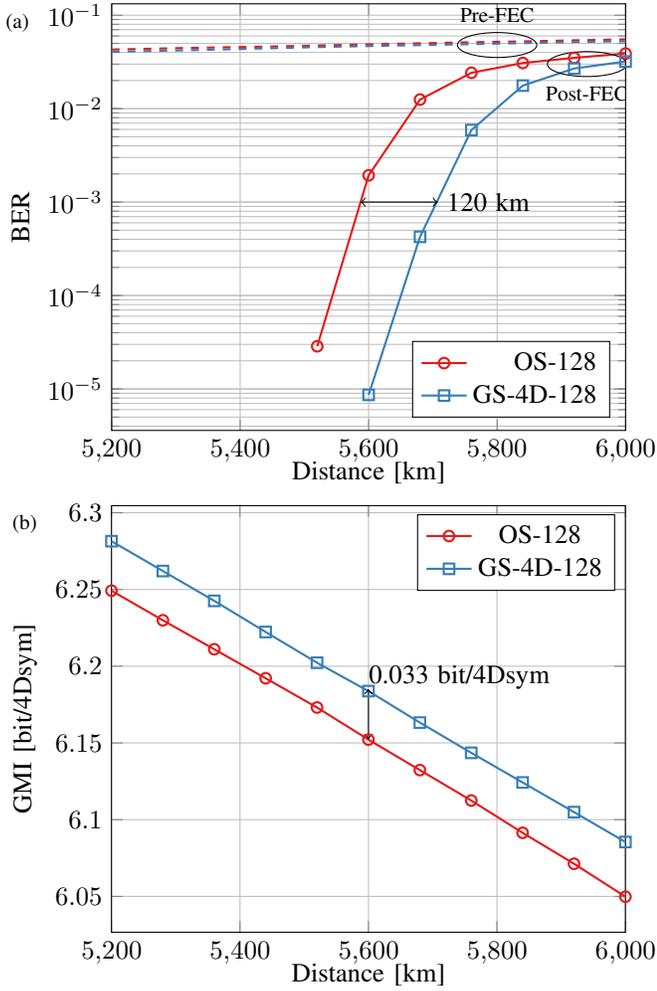
\begin{figure}
    \centering

\begin{tikzpicture}
    \begin{groupplot}[
            group style = {group size = 1 by 2},
            width=0.95\columnwidth,
            xmin=5200,
            xmax=6000,
            grid=both,
        ]
        \nextgroupplot[
            xlabel={Distance [km]},
            ylabel={BER},
            ymode=log,
            grid=both,
            legend style={
            cells={anchor=east},
            legend pos=south east,
            xlabel shift=-7pt,
            },]
        \coordinate (a) at (yticklabel cs:1);
        
        \addplot[Set1-A,thick,mark=none,dashed,forget plot] table[x=L,y=preBER] {data/OS128/os_res.txt};
        \addplot[Set1-A,thick,mark=o] table[x=L,y=postBER] {data/OS128/os_res.txt};
        \addlegendentry{OS-128}

        \addplot[Set1-B,thick,dashed,forget plot] table[x=L,y=preBER] {data/OS128/bin_res.txt};
        \addplot[Set1-B,thick,mark=square] table[x=L,y=postBER] {data/OS128/bin_res.txt};
        \addlegendentry{GS-4D-128}
        
        \node[draw, ellipse, minimum width=30pt,label=above:{\footnotesize Pre-FEC}] at (axis cs:5800,0.048) {};
        \node[draw, ellipse, minimum width=30pt,label=below:{\footnotesize Post-FEC}] at (axis cs:5940,0.03) {};

        \draw[<->] (axis cs:5587.5,0.001) -- (axis cs:5707.5,0.001) node[pos=1, anchor=west] {120 km};

        \nextgroupplot[
            xlabel={Distance [km]},
            ylabel={GMI [bit/4Dsym]},
            xlabel shift=-7pt,
            ]
        \coordinate (b) at (yticklabel cs:1);
        \addplot[Set1-A,thick,mark=o] table[x=L,y=GMI] {data/OS128/os_res.txt};
        \addlegendentry{OS-128}

        \addplot[Set1-B,thick,mark=square] table[x=L,y=GMI] {data/OS128/bin_res.txt};
        \addlegendentry{GS-4D-128}

        \draw[<->] (axis cs:5600,6.1525) -- (axis cs:5600,6.185) node[pos=1, anchor=south west, inner sep=0] {0.033 bit/4Dsym};
    \end{groupplot}
    \node[anchor=north east] at (a) {\footnotesize (a)};
    \node[anchor=north east] at (b) {\footnotesize (b)};
\end{tikzpicture}

     \caption{Comparison between the OS-128 constellation from \cite{chen2020analysis} and a 128-ary constellation designed with the method shown in this work on a simulated fibre system. (a) shows the pre- and post-FEC BER for the $R=4/5$ DVB-S2 LDPC and (b) shows the GMI vs the transmitted distance.}
    \label{fig:longhaul_results}
\end{figure}

The performance of the resulting \ac{GS-4D} constellation is shown in Fig.~\ref{fig:longhaul_results}. We can see that the additional freedom from optimising all points in the orthant individually allowed our constellation to outperform the constellation obtained using the approach in \cite{chen2020analysis}, albeit by a small margin (0.033 bit/4Dsym) or 120~km; a similar observation to that made by the authors in \cite{chen2020analysis}. The results are shown in terms of both \ac{BER} and \ac{GMI}. For the \ac{BER} a post-\ac{FEC} curve is shown where the $R=4/5$ \ac{LDPC} from \cite{Standard2014DVBS2X} is applied.

\begin{figure}
    \centering
\begin{tikzpicture}
        \begin{groupplot}[
            group style = {group size = 1 by 2},
            width=0.93\columnwidth,
            grid=both,
            legend style={font=\footnotesize},
            colormap name=isoL-10,
            cycle list={[ of colormap]},
            xmin=0,xmax=10000,
            extra x ticks={2000,4000,6000},
            log x ticks with fixed point,
            extra x tick style={log identify minor tick positions=false},
            ]
        
            \nextgroupplot[
                xmin=800,
                xmode=log,
                ymode=log,
                legend columns=3,
                reverse legend,
                transpose legend,
                legend style={at={(0.5,0.03)},anchor=south},
                xlabel={Distance [km]},
                ylabel={BER},
                thick, xlabel shift=-7pt]
        \coordinate (a) at (yticklabel cs:1);

            \addlegendimage{dashed}
            \addlegendentry{Pre-FEC BER}
            \addlegendimage{}
            \addlegendentry{Post-FEC BER}

            \addplot+[dashed,forget plot] table[x=L,y=preBER] {data/osrange/bind32_res.txt};
            \addplot table[x=L,y=postBER] {data/osrange/bind32_res.txt};
            \addlegendentry{32}

            \addplot+[dashed,forget plot] table[x=L,y=preBER] {data/osrange/bin64_res_a.txt};
            \addplot table[x=L,y=postBER] {data/osrange/bin64_res_a.txt};
            \addlegendentry{64}

            \addplot+[dashed,forget plot] table[x=L,y=preBER] {data/osrange/bin128_res.txt};
            \addplot table[x=L,y=postBER] {data/osrange/bin128_res.txt};
            \addlegendentry{128}   

            \addplot+[dashed,forget plot] table[x=L,y=preBER] {data/osrange/bin256_res.txt};
            \addplot table[x=L,y=postBER] {data/osrange/bin256_res.txt};
            \addlegendentry{256}  

            \addplot+[dashed,forget plot] table[x=L,y=preBER] {data/osrange/bin512_res.txt};
            \addplot table[x=L,y=postBER] {data/osrange/bin512_res.txt};
            \addlegendentry{512}  

            \addplot+[dashed,forget plot] table[x=L,y=preBER] {data/osrange/bin1024_res.txt};
            \addplot table[x=L,y=postBER] {data/osrange/bin1024_res.txt};
            \addlegendentry{1024}  

            \addplot+[dashed,forget plot] table[x=L,y=preBER] {data/osrange/bin2048_res.txt};
            \addplot table[x=L,y=postBER] {data/osrange/bin2048_res.txt};
            \addlegendentry{2048}  

            \addplot+[dashed,forget plot] table[x=L,y=preBER] {data/osrange/bin4096_res.txt};
            \addplot table[x=L,y=postBER] {data/osrange/bin4096_res.txt};
            \addlegendentry{4096}  

            \addplot+[dashed,forget plot] table[x=L,y=preBER] {data/osrange/bin8192_res.txt};
            \addplot table[x=L,y=postBER] {data/osrange/bin8192_res.txt};
            \addlegendentry{8192}  

            \nextgroupplot[
                xmin=800,
                xmode=log,
                ytick distance=1.6,
                legend columns=3,
                reverse legend,
                transpose legend,
                legend style={at={(0.5,0)},anchor=south,},
                ymin=0,ymax=12,
                xlabel={Distance [km]},
                ylabel={AIR [bit/4Dsym]},
                thick,xlabel shift=-7pt]
        \coordinate (b) at (yticklabel cs:1);

            \addlegendimage{dashed}
            \addlegendentry{GMI}
            \addlegendimage{}
            \addlegendentry{Post-FEC $R^*$}

            \addplot+[dashed,forget plot] table[x=L,y=GMI] {data/osrange/bind32_res.txt};
            \addplot table[x=L,y expr=0.8*5*(1+\thisrow{postBER}*log2(\thisrow{postBER})+(1-\thisrow{postBER})*log2(1-\thisrow{postBER}))] {data/osrange/bind32_res.txt};
            \addlegendentry{32}

            \addplot+[dashed,forget plot] table[x=L,y=GMI] {data/osrange/bin64_res_a.txt};
            \addplot table[x=L,y expr=0.8*6*(1+\thisrow{postBER}*log2(\thisrow{postBER})+(1-\thisrow{postBER})*log2(1-\thisrow{postBER}))] {data/osrange/bin64_res_a.txt};
            \addlegendentry{64}

            \addplot+[dashed,forget plot] table[x=L,y=GMI] {data/osrange/bin128_res.txt};
            \addplot table[x=L,y expr=0.8*7*(1+\thisrow{postBER}*log2(\thisrow{postBER})+(1-\thisrow{postBER})*log2(1-\thisrow{postBER}))] {data/osrange/bin128_res.txt};
            \addlegendentry{128}

            \addplot+[dashed,forget plot] table[x=L,y=GMI] {data/osrange/bin256_res.txt};
            \addplot table[x=L,y expr=0.8*8*(1+\thisrow{postBER}*log2(\thisrow{postBER})+(1-\thisrow{postBER})*log2(1-\thisrow{postBER}))] {data/osrange/bin256_res.txt};
            \addlegendentry{256}

            \addplot+[dashed,forget plot] table[x=L,y=GMI] {data/osrange/bin512_res.txt};
            \addplot table[x=L,y expr=0.8*9*(1+\thisrow{postBER}*log2(\thisrow{postBER})+(1-\thisrow{postBER})*log2(1-\thisrow{postBER}))] {data/osrange/bin512_res.txt};
            \addlegendentry{512}

            \addplot+[dashed,forget plot] table[x=L,y=GMI] {data/osrange/bin1024_res.txt};
            \addplot table[x=L,y expr=0.8*10*(1+\thisrow{postBER}*log2(\thisrow{postBER})+(1-\thisrow{postBER})*log2(1-\thisrow{postBER}))] {data/osrange/bin1024_res.txt};
            \addlegendentry{1024}

            \addplot+[dashed,forget plot] table[x=L,y=GMI] {data/osrange/bin2048_res.txt};
            \addplot table[x=L,y expr=0.8*11*(1+\thisrow{postBER}*log2(\thisrow{postBER})+(1-\thisrow{postBER})*log2(1-\thisrow{postBER}))] {data/osrange/bin2048_res.txt};
            \addlegendentry{2048}

            \addplot+[dashed,forget plot] table[x=L,y=GMI] {data/osrange/bin4096_res.txt};
            \addplot table[x=L,y expr=0.8*12*(1+\thisrow{postBER}*log2(\thisrow{postBER})+(1-\thisrow{postBER})*log2(1-\thisrow{postBER}))] {data/osrange/bin4096_res.txt};
            \addlegendentry{4096}

            \addplot+[dashed,forget plot] table[x=L,y=GMI] {data/osrange/bin8192_res.txt};
            \addplot table[x=L,y expr=0.8*13*(1+\thisrow{postBER}*log2(\thisrow{postBER})+(1-\thisrow{postBER})*log2(1-\thisrow{postBER}))] {data/osrange/bin8192_res.txt};
            \addlegendentry{8192}

    \end{groupplot}
    \node[anchor=north east] at (a) {\footnotesize (a)};
    \node[anchor=north east] at (b) {\footnotesize (b)};
\end{tikzpicture}

             \caption{Comparisons of the GS-4D constellations shaped for the \ac{AWGN} channel. (a) shows the pre- and post-FEC BER for the $R=4/5$ DVB-S2 LDPC and (b) shows the GMI and the post-FEC rate vs the transmitted distance. Numbers in the legend indicate the constellation order $M$.}
    \label{fig:os_range}
\end{figure}
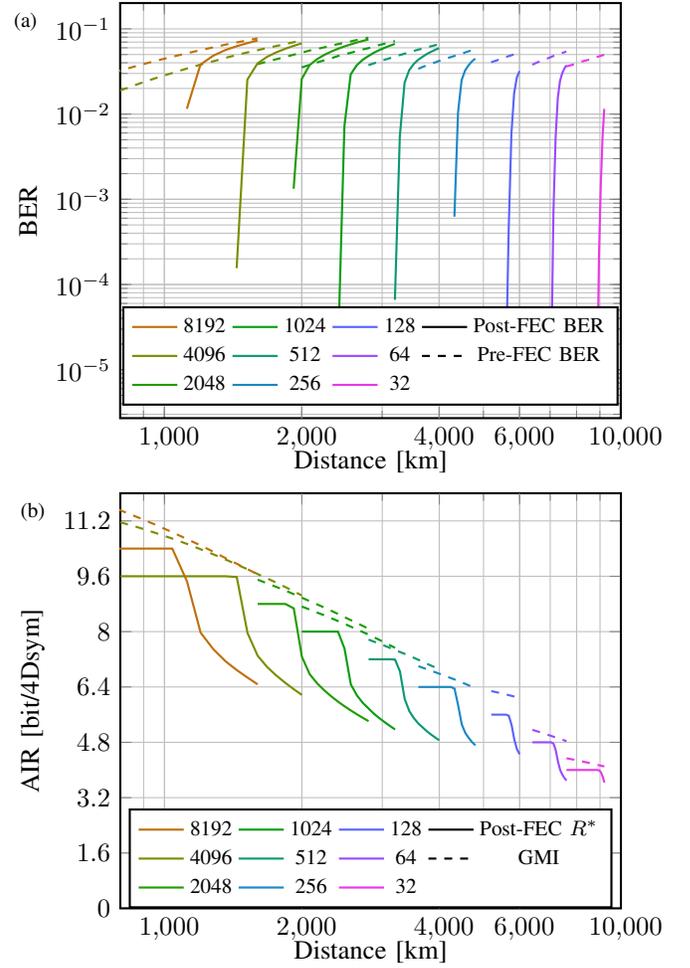

However, because the optimisation in this work does not rely on higher-order symmetry to reduce the number of optimisation variables and it is significantly more computationally efficient to optimise, we can now optimise constellations with a large number of constellation points. In Fig.~\ref{fig:os_range}, the results using the same simulation system for a greater number of constellation points are shown. The orthant symmetry is kept as the lower \acp{SNR} does not result in worse performance. The results are shown in terms of both \ac{BER} and \ac{GMI}; for the \ac{BER} a post-\ac{FEC} curve is shown for a $R=4/5$ \ac{LDPC} code from \cite{Standard2014DVBS2X}. 
The performance of the \ac{LDPC}-coded scheme is shown in terms of $R^*$ from Eq.~\eqref{eq:R_star}.
We can see that for constellations with a number of points ranging from 32 to 8196 in 4 dimensions, we have successfully designed constellations targeting the $R=4/5$ rate. The $R^*$ saturates at $m R$ for shorter distances until it reaches the \ac{GMI}. Additionally, the 4096-ary constellation from this optimisation is used in Sec.~\ref{sec:resultnonlin}, where it is compared against other constellations with the same number of bits per dimension in Fig.~\ref{fig:compare64}.

\section{Results for the Nonlinear Fibre Channel}
\label{sec:resultnonlin}

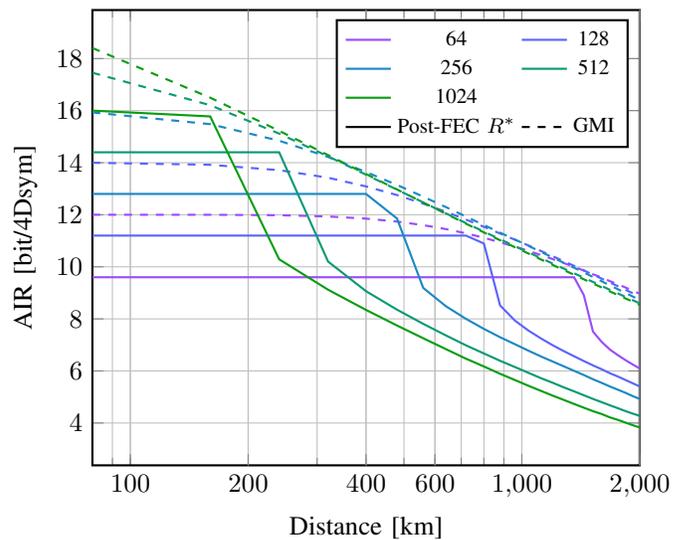
\begin{figure}
    \centering
\begin{tikzpicture}
    \begin{axis}[
        width=\columnwidth,
        legend style={font=\footnotesize},
        colormap name=isoL-10,
        cycle list={[ of colormap]},
        cycle list shift=1,
        xmode=log,
        grid=both,
        ytick distance=2,
        xmin=80,xmax=2000,
        extra x ticks={200,400,600,2000},
        log ticks with fixed point,
        extra x tick style={log identify minor tick positions=false},
        xlabel={Distance [km]},
        ylabel={AIR [bit/4Dsym]},
        legend columns=2,
        thick]
    ]
    
        \addplot+[dashed,forget plot] table[x=L,y expr=2*\thisrow{GMI}]{data/nonlin2D/2Dnolin64_res.txt};
        \addplot table[x=L,y expr=2*6*0.8*(1+\thisrow{postBER}*log2(\thisrow{postBER})+(1-\thisrow{postBER})*log2(1-\thisrow{postBER}))]{data/nonlin2D/2Dnolin64_res.txt};
        \addlegendentry{64}

        \addplot+[dashed,forget plot] table[x=L,y expr=2*\thisrow{GMI}]{data/nonlin2D/2Dnolin128_res.txt};
        \addplot table[x=L,y expr=2*7*0.8*(1+\thisrow{postBER}*log2(\thisrow{postBER})+(1-\thisrow{postBER})*log2(1-\thisrow{postBER}))]{data/nonlin2D/2Dnolin128_res.txt};
        \addlegendentry{128}

        \addplot+[dashed,forget plot] table[x=L,y expr=2*\thisrow{GMI}]{data/nonlin2D/2Dnolin256_res.txt};
        \addplot table[x=L,y expr=2*8*0.8*(1+\thisrow{postBER}*log2(\thisrow{postBER})+(1-\thisrow{postBER})*log2(1-\thisrow{postBER}))]{data/nonlin2D/2Dnolin256_res.txt};
        \addlegendentry{256}

        \addplot+[dashed,forget plot] table[x=L,y expr=2*\thisrow{GMI}]{data/nonlin2D/2Dnolin512_res.txt};
        \addplot table[x=L,y expr=2*9*0.8*(1+\thisrow{postBER}*log2(\thisrow{postBER})+(1-\thisrow{postBER})*log2(1-\thisrow{postBER}))]{data/nonlin2D/2Dnolin512_res.txt};
        \addlegendentry{512}

        \addplot+[dashed,forget plot] table[x=L,y expr=2*\thisrow{GMI}]{data/nonlin2D/2Dnolin1024_res.txt};
        \addplot table[x=L,y expr=2*10*0.8*(1+\thisrow{postBER}*log2(\thisrow{postBER})+(1-\thisrow{postBER})*log2(1-\thisrow{postBER}))]{data/nonlin2D/2Dnolin1024_res.txt};
        \addlegendentry{1024}

        \addlegendimage{empty legend}\addlegendentry{}
            \addlegendimage{}
            \addlegendentry{Post-FEC $R^*$}
            \addlegendimage{dashed}
            \addlegendentry{GMI}

    \end{axis}
\end{tikzpicture}     \caption{Comparison of the GS-NL-2D constellations, which are shaped to trade off linear and nonlinear shaping gain, designed for $R=4/5$ code. The GMI (dashed lines) and the post-\ac{FEC} rate $R^*$ are plotted against the transmitted distance. Numbers in the legend indicate the constellation order.}
    \label{fig:nonlin2D}
\end{figure}

The results of the shaping of 2D constellations, shaped for the nonlinear fibre channel, as described in Sec.~\ref{sec:channel_model}, are shown in Fig.~\ref{fig:nonlin2D}.  The results shown were calculated using the same \ac{SSFM} simulation setup as described in Sec.~\ref{sec:shape4d}. The constellations were designed to maximise the trade-off between shaping gain and nonlinear distortion. This was achieved by numerical optimisation, again using the trust-region algorithm, using Eq.~\eqref{eq:nonlin_objective} as the objective function. The eta ratio $c$ was estimated to be around 0.4, established by fitting the performance difference between \ac{QPSK} transmission and transmission where the sequence was drawn from a normal distribution as the reference distribution. These results present a range of constellations for a fixed eta ratio since the performance increase is not guaranteed if this changes. For a given system, where the length, number of amplifiers and spectral neighbours are fixed, these constellations show an improvement in performance. The lower cardinality constellations perform better for longer distances and saturate for shorter distances. The \ac{GMI} saturates to $m$, the $R^*$ saturates to $mR$. At the target reach, the $R^*$ gets close to the GMI, this is the SNR for which the constellations were optimised.

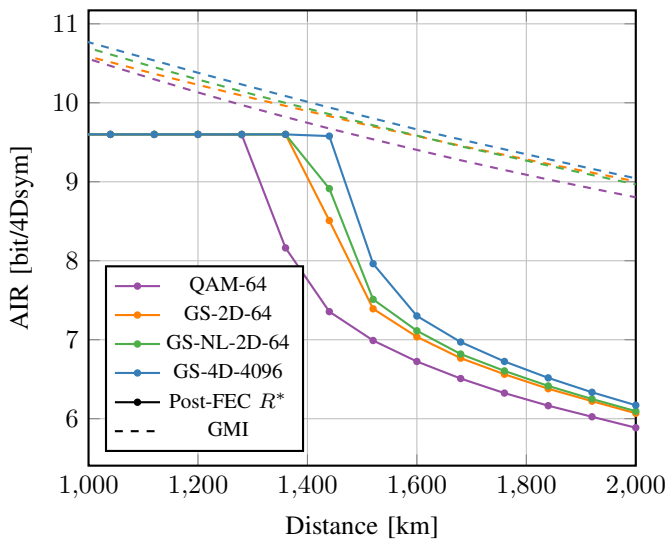
\begin{figure}
    \centering
\begin{tikzpicture}
    \begin{axis}[
        width=\columnwidth,
        grid=both,
        xlabel={Distance [km]},
        ylabel={AIR [bit/4Dsym]},
        xmin=1000,xmax=2000,
        legend style={font=\footnotesize},
        legend pos={south west},
        thick,
    ]
        
        \addplot[Set1-D,dashed,forget plot] table[x=L,y expr=2*\thisrow{GMI}]{data/R45square64_res.txt};
        \addplot[Set1-D,mark=*,mark size=1pt] table[x=L,y expr=2*6*0.8*(1+\thisrow{postBER}*log2(\thisrow{postBER})+(1-\thisrow{postBER})*log2(1-\thisrow{postBER}))]{data/R45square64_res.txt};
        \addlegendentry{QAM-64}

        \addplot[Set1-E,dashed,forget plot] table[x=L,y expr=2*\thisrow{GMI}]{data/R45op15dB_64_res.txt};
        \addplot[Set1-E,mark=*,mark size=1pt] table[x=L,y expr=2*6*0.8*(1+\thisrow{postBER}*log2(\thisrow{postBER})+(1-\thisrow{postBER})*log2(1-\thisrow{postBER}))]{data/R45op15dB_64_res.txt};
        \addlegendentry{GS-2D-64}

        \addplot[Set1-C,dashed,forget plot] table[x=L,y expr=2*\thisrow{GMI}]{data/nonlin2D/2Dnolin64_res.txt};
        \addplot[Set1-C,mark=*,mark size=1pt] table[x=L,y expr=2*6*0.8*(1+\thisrow{postBER}*log2(\thisrow{postBER})+(1-\thisrow{postBER})*log2(1-\thisrow{postBER}))]{data/nonlin2D/2Dnolin64_res.txt};
        \addlegendentry{GS-NL-2D-64}

        \addplot[Set1-B,dashed,forget plot] table[x=L,y=GMI] {data/osrange/bin4096_res.txt};
        \addplot[Set1-B,mark=*,mark size=1pt] table[x=L,y expr=0.8*12*(1+\thisrow{postBER}*log2(\thisrow{postBER})+(1-\thisrow{postBER})*log2(1-\thisrow{postBER}))] {data/osrange/bin4096_res.txt};
        \addlegendentry{GS-4D-4096}

            \addlegendimage{mark=*,mark size=1pt}
            \addlegendentry{Post-FEC $R^*$}
            \addlegendimage{dashed}
            \addlegendentry{GMI}

    \end{axis}
\end{tikzpicture}     \caption{Comparison of constellation formats with 64 points per 2D. The GMI and the post-\ac{FEC} rate $R^*$ are plotted versus the transmitted distance.}
    \label{fig:compare64}
\end{figure}

The comparison of the different strategies investigated in this paper is shown in Fig.~\ref{fig:compare64} evaluated using the \ac{SSFM} as described previously. This figure shows the performance of 4 different formats, namely the regular square \ac{QAM}, 2D and 4D constellations optimised for the \ac{AWGN} channel and a 2D constellation optimised with the same number of bits per 4 dimensions, in terms of the trade-off between the linear and nonlinear shaping gain. All curves saturate for shorter transmitted distances, the \ac{GMI} at $m=12$ bit/4D and the $R^*$ at $mR=9.6$ bit/4D.  The first result shown is regular \ac{QAM}. Following this, the 2D constellation shaped for AWGN is presented, which exhibits better performance. The 2D constellation shaped for the linear/nonlinear trade-off has a slightly better performance. The 4D constellation for \ac{AWGN} has the best performance in terms of \ac{AIR}. it should be noted that although it does not show the best performance in terms of SNR,  it results in a higher \ac{AIR} or throughput, in terms of GMI, through the resultant shaping gain. The nonlinear performance metrics presented in this paper are only valid for 2D constellations and, therefore, applying these metrics to the 4D constellation yielded no further performance improvements and the reason why is the subject of current research\cite{Liga2020Extending}. It can be seen, however, that all the shaped constellations outperform the regular \ac{QAM}, after a 1520-km transmission; the 4D constellation by 0.27~bit/4D, the constellation shaped for the nonlinear channel with 0.18 bit/4D and the 2D constellation shaped for \ac{AWGN} by 0.16 bit/4D in \ac{GMI}. The $R^*$ shows that the GS-4D-4096 has a reach increase of 2 spans, whereas the GS-2D and the GS-NL-2D show a reach increase of more than 1 span, where after 1440~km transmission, the GS-NL-2D outperforms the GS-2D by 0.4~bit/4D. 

\begin{figure}
    \centering
\begin{tikzpicture}
    \begin{axis}[
        width=\columnwidth,
        grid=both,
        xlabel={Distance [km]},
        ylabel={SNR [dB]},
        xtick distance=40,
        xmin=1360,xmax=1520,
        legend columns=3,
        legend style={font=\footnotesize},
        thick,
    ]
    
    \addplot[Set1-A] table[x=L,y=QPSK]{data/gnspsk.txt};
    \addlegendentry{QPSK}

    \addplot[Set1-D] table[x=L,y=SNR]{data/R45square64_res.txt};
    \addlegendentry{QAM-64}

    \addplot[Set1-E] table[x=L,y=SNR]{data/R45op15dB_64_res.txt};
    \addlegendentry{GS-2D-64}

    \addplot[Set1-C] table[x=L,y=SNR]{data/nonlin2D/2Dnolin64_res.txt};
    \addlegendentry{GS-NL-2D-64}

    \addplot[Set1-B] table[x=L,y=SNR] {data/osrange/bin4096_res.txt};
    \addlegendentry{GS-4D-4096}

    \addplot[Set1-I] table[x=L,y=GN]{data/gnspsk.txt};
    \addlegendentry{Gaussian}

    \end{axis}
\end{tikzpicture}     \caption{Comparison of transmission performance for constellations with 64 points per 2D. The achievable SNR for QPSK and Gaussian modulation is shown for reference.}
    \label{fig:compareSNR}
\end{figure}
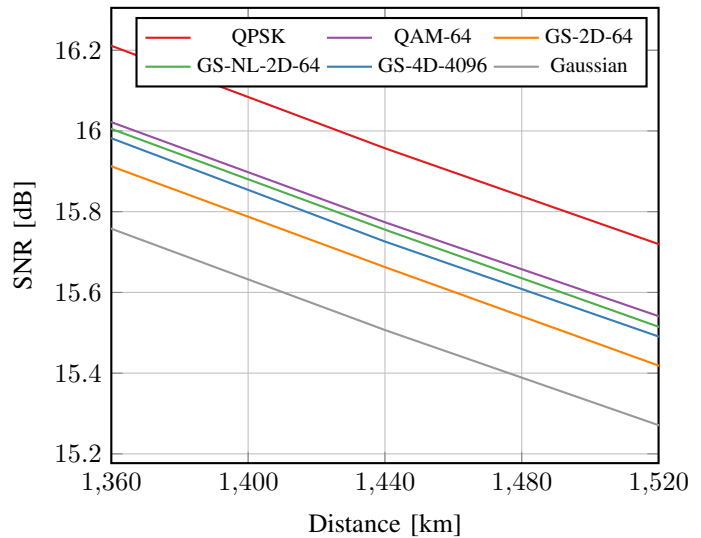

The performance in terms of \ac{SNR} for each constellation, is shown in Fig.~\ref{fig:compareSNR}. With the SNR results, we can explain some of the performance differences in Fig.~\ref{fig:compare64}. The results also show the SNR for \ac{QPSK} and Gaussian modulation for comparison, as these represent the reasonable best and worst nonlinear channel responses, respectively.
For the reference case of Gaussian modulation, the transmitted sequence was generated with samples taken from a normal distribution, treated as symbols.

The effect of each constellation format on the nonlinear performance can be observed. As expected, the square QAM performed the best and the GS-2D constellation shaped for the AWGN channel performs the worst with $0.1$ dB between them.
However, the GS-2D constellation still outperformed the square QAM by at least one span or $0.1$ bit higher GMI.

It can be seen that the GS-4D constellation has the best performance overall. The constellation exhibits very good nonlinear performance despite not being optimised for nonlinearity, increasing the SNR at optimum launch power. Together with excellent shaping gain from the additional degrees of freedom, this 4D AWGN-optimised constellation was the best performer in our selection.

For the case of the GS-NL-2D constellation, shaped using Eq.~\eqref{eq:nonlin_objective} as optimisation function, the nonlinear SNR penalty was reduced to a value close to that of the square 64-QAM whilst retaining shaping gain. The loss in shaping gain is more than offset by the improvement of the nonlinear channel response. 

 The drawback of the 4D constellation is that it requires a 4D demapper, meaning that instead of 2$\times$64 2D distances, 4096 4D distances must be evaluated; a factor of 64 increase in the number of points and a factor of two again for the doubling in the number of the dimensions per constellation point. The choice of constellation thus becomes a trade-off between performance and complexity. 

\begin{figure}
    \centering
\begin{tikzpicture}
\begin{axis}[
        width=\columnwidth,
        grid=both,
        xlabel={Launch Power per Channel [dBm]},
        ylabel={AIR [bit/4Dsym]},
        legend columns=2,
        transpose legend,
        legend style={font=\footnotesize, anchor=south, at={(axis cs:-0.5,6)}},
        thick,]

\addplot[Set1-D, dashed, forget plot] table[x=Plaunch,y=GMI] {data/1440km/1440square64_res.txt};
\addplot[Set1-D, mark=*, mark size=1pt] table[x=Plaunch,y=Rstar] {data/1440km/1440square64_res.txt};
\addlegendentry{QAM-64}

\addplot[Set1-E, dashed, forget plot] table[x=Plaunch,y=GMI] {data/1440km/1440R45op15dB_64_res_res.txt};
\addplot[Set1-E, mark=*, mark size=1pt] table[x=Plaunch,y=Rstar] {data/1440km/1440R45op15dB_64_res_res.txt};
\addlegendentry{GS-2D-64}

\addplot[Set1-C, dashed, forget plot] table[x=Plaunch,y=GMI] {data/1440km/1440nolin64_res.txt};
\addplot[Set1-C, mark=*, mark size=1pt] table[x=Plaunch,y=Rstar] {data/1440km/1440nolin64_res.txt};
\addlegendentry{GS-NL-2D-64}

\addplot[Set1-B, dashed, forget plot] table[x=Plaunch,y=GMI] {data/1440km/1440gs4_4096_res.txt};
\addplot[Set1-B, mark=*, mark size=1pt] table[x=Plaunch,y=Rstar] {data/1440km/1440gs4_4096_res.txt};
\addlegendentry{GS-4D-4096}

\addlegendimage{dashed}
\addlegendentry{GMI}

\addlegendimage{mark=*, mark size=1pt}
\addlegendentry{$R^*$}

\end{axis}
\end{tikzpicture}     \caption{Comparison of achievable rates for constellations with 64 points per 2D versus launch power at a transmission distance of 1440~km.}
    \label{fig:compareLaunchPower}
\end{figure}
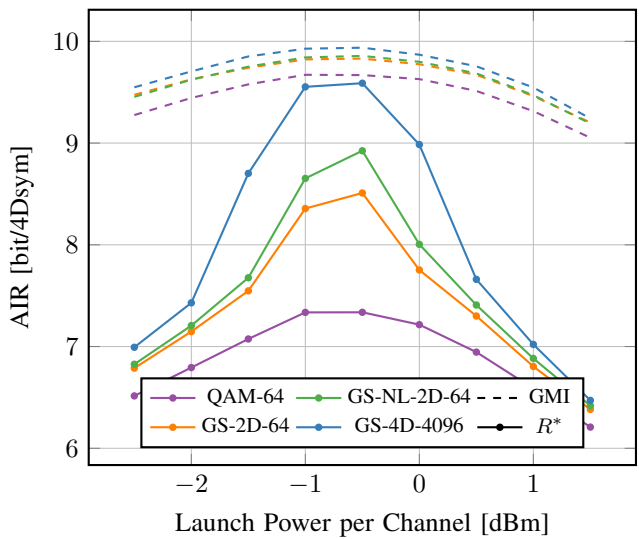

In Fig.~\ref{fig:compareLaunchPower}, the geometrically shaped constellations are compared at the transmission distance of 1440~km. Here the launched power is swept and it can be seen that the optimum launch power is close to 0.5~dBm per channel used previously. The GS-NL-2D constellation outperforms the GS-2D even when the GMI is lower, suggesting that resulting LLRs result in better LDPC performance, although more research is necessary to reach a conclusion on that.

The choice between the GS-2D and GS-NL-2D would then depend on the system's operational consideration. When a system is expected to be fully loaded, the lesser impact on neighbouring channels of the GS-NL-2D may be attractive. When the system is expected to operate more sparsely, with lower nonlinear interaction between channels, the improved performance through maximising shaping gain with GS-2D can prove to be more beneficial. Ideally, this change should be applied in an adaptive manner, in response to the change in system conditions such as demand, transmission distances and the number of occupied wavelength channels.

\section{Computational Complexity}

Whilst we optimise the geometry of the constellation, all constellation points influence each other and the performance of the constellation as a whole is measured by a single scalar metric, i.e., the \ac{MI} or \ac{GMI}. That means the objective function can be seen as a function that takes in an $M \times 2N$ matrix and outputs a scalar value. For some optimisation techniques, it is convenient to re-shape the input as a vector, this way the gradient and Hessian are defined.

The slowest operation in the calculation is the evaluation of Eq.~\eqref{eq:hfunc} and the summation for \ac{GHQ}. Fortunately, calculating the gradient in addition to the function does not require additional evaluations of Eq.~\eqref{eq:hfunc}, just an additional summation for the \ac{GHQ}. This makes the evaluation of the function and gradient approximately two times slower than just evaluating the function. To investigate the scaling of the computational complexity, we can simply count how many times Eq.~\eqref{eq:hfunc} is evaluated. For the \ac{MI}, it is calculated for every $i$ and every $j$ multiplied by the number of points used for the \ac{GHQ}. Therefore, if we have $M$ constellation points and $L$ \ac{GHQ} points per dimension, the computational complexity scales with $\mathcal{O}(M^2\times L^{2N})$. For \ac{GMI}, the complexity scales with $\mathcal{O}\big(M^2\times L^{2N}+ \log_2(M)\times M\times\frac{M}{2}\times L^{2N}\big)$. When increasing the number of dimensions, the number of constellation points must grow exponentially with the number of dimensions if the number of bits per dimension is to remain constant. If we define $\hat{m} = \frac{\log_2(M)}{N}$ as the number of bits per 2 dimensions, then the \ac{GMI} scales with $\mathcal{O}\big((2^{\hat{m}2N}+ \hat{m}N\times 2^{\hat{m}4N})\times L^{2N}\big)$.

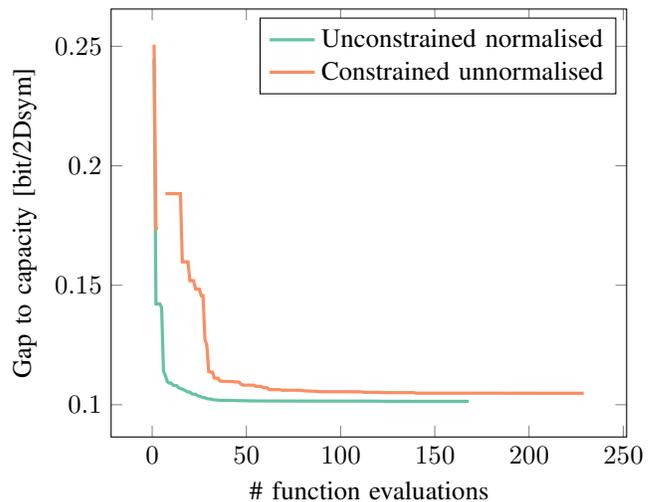
\begin{figure}
    \centering
\begin{tikzpicture}
     \begin{axis}[
     xlabel={\# function evaluations},
     ylabel={Gap to capacity [bit/2Dsym]},
     unbounded coords=jump,
     ]
            \addplot[Set2-A,very thick] table[x index=2, y expr=log2(1+10^(21/10))+\thisrowno{1}] {data/opt_unconstrained.txt};
            \addlegendentry{Unconstrained normalised}
            \addplot[Set2-B,very thick] table[x index=3, y expr=log2(1+10^(21/10))+\thisrowno{1}] {data/opt_constrained.txt};
            \addlegendentry{Constrained unnormalised}
     \end{axis}
\end{tikzpicture}
    \caption{Comparison of convergence speed between an unconstrained optimisation where the gradient includes normalisation and constrained optimisation where the gradient does not include the normalisation. Both optimising 1024 QAM for AWGN starting from APSK.}
    \label{fig:convergence}
\end{figure}

Typically, the gradient of the function can be calculated using the finite difference method, which allows to obtain numerically the gradient for optimisation. For a single input and a single output, \cite[Ch.~25]{Abramowitz1965Handbook} provides an expression for the calculation of finite differences. The derivative of a function $g$ at a point $x$ is approximated via the finite-difference method $(g(x+h) -g(x))/h$ for some small value of $h$. When the function takes multiple inputs, the procedure is repeated for every input variable.
It is important to note that this method requires an extra function evaluation for every component of the input space. This will quickly become unfeasible; for example, optimising a 4D 4096-QAM constellation requires 4$\times$4096 $=$ 16384 evaluations of the objective function. 

Our method, combining the normalisation with the objective function of the optimisation, allows us to use unconstrained optimisation methods. This is superior to using the \ac{MI} or \ac{GMI} with respect to an input constellation directly as an objective function and prevents the constellation amplitude from growing using a power constraint in the optimisation. This allows us to use unconstrained quasi-Newton optimisation algorithms, as shown in Fig.~\ref{fig:convergence}.
 In this example, 1024 QAM was optimised for the \ac{AWGN} channel starting from an \ac{APSK} constellation described in \cite{Meric2015Approaching}. 
 Another benefit of combining the gradient and channel constraint function this way is that it allows the channel function to include other channel effects, for example residual phase noise or device nonlinearities, beyond optical transmission systems.

\section{Conclusions}
\label{sec:conclusions}

This paper presents a novel and fast computation method for optimising geometrically shaped signal constellations with very high cardinality.

We proposed the use of an analytically derived gradient, instead of the conventionally-used finite-difference method, significantly reducing the complexity of numerical optimisation. Additionally, we combined the gradient of the mutual information with the Jacobian of the channel constraints, translating the problem into an unconstrained optimisation problem. We have successfully applied this technique to both linear and nonlinear optical communication systems.

We applied the proposed algorithms to show shaping gain for high cardinality 2- and 4- dimensional constellations, to generate geometrically shaped constellations, with up to 8192 constellation points. An 8192-ary constellation achieving a GMI with a gap of just 0.06 bit/2D to the AWGN channel capacity was designed, with the gains in mutual information verified using a simulation of a nonlinear optical fibre communication system, exhibiting shaping gains in both the linear and nonlinear regimes.

Although the focus of this paper is on optical communications, the same approach can be applied  more generally to design and optimise a wide range of other communication systems operating in the linear and nonlinear regimes.

\appendices
\section{Derivation of the Gradient of the MI}
\label{ap:mi_derivation}

The gradient of the \ac{MI} is a vector partial derivative
\begin{equation}
    \nabla I = [\ddx{\vect{x}_1}I, \ddx{\vect{x}_2}I,\ldots,\ddx{\vect{x}_M}I ] \,.
\end{equation}%

To derive the gradient more easily we consider the different parts of the equation. First, we can define a function $g_i(\vect{z},\mat{x}_\mathcal{J})$ for the MI and set $\setJi=\{j\in \mathcal{J}; j\neq i\}$, which depends on $h(\vect{z},\vect{x}_i,\vect{x}_j)$
\begin{align}
    g_i(\vect{z},\mat{x}_\mathcal{J}) &\triangleq \log_2\left( \sum_{j\in\mathcal{J}} h(\vect{z},\vect{x}_i,\vect{x}_j) \right) \nonumber\\
    &=\log_2\left( 1 + \sum_{j\in\setJi} h(\vect{z},\vect{x}_i,\vect{x}_j)\right) \label{eq:g_subfunction} \\
    h(\vect{z},\vect{x}_i,\vect{x}_j) &\triangleq \exp \left(\frac{||\vect{x}_i-\vect{x}_j||^2 + 2 \langle \vect{z}, (\vect{x}_i-\vect{x}_j)\rangle}{-\sigma_z^2}\right) \,. \label{eq:h_subfunction}
\end{align}
For these building blocks, the partial derivative can be easily obtained. 

The partial derivatives in the gradient can be derived using the Leibniz and the chain rule:
\begin{align}
\begin{aligned}
    \ddx{\vect{x}_n}I = -&\frac{1}{M} \E\ddx{\vect{x}_n}g_n(\vect{z},\mat{x}_\mathcal{J})\dz \\ -&\frac{1}{M}\sum_{i\in\setJn}  \E \ddx{\vect{x}_{n}}g_i(\vect{z},\mat{x}_\mathcal{J})\dz \end{aligned} \,. \label{eq:mi_derrivative}
\end{align}
The sum over $i$ is already split into the cases $i=n$ and $i\neq n$, which simplifies the next steps.

To evaluate this gradient, we start with the partial derivative of Eq.~\eqref{eq:g_subfunction}
\begin{equation}
 \ddx{\vect{x}_n}g_i(\vect{z},\mat{x}_\mathcal{J}) =
 \begin{cases}
    \frac{\sum_{j\in\setJn} \ddx{\vect{x}_n}h(\vect{z},\vect{x}_n,\vect{x}_j) }{\log(2)\sum_{j\in\mathcal{J}} h(\vect{z},\vect{x}_n,\vect{x}_j) } &\text{for } i = n\\
    \frac{\ddx{\vect{x}_n}h(\vect{z},\vect{x}_i,\vect{x}_n) }{\log(2)\sum_{j\in\mathcal{J}}^M h(\vect{z},\vect{x}_i,\vect{x}_j) } &\text{for } i \neq n
 \end{cases}\label{eq:partial_g} \,,
\end{equation} 
with which we can then arrive to partial derivative of Eq.~\eqref{eq:h_subfunction} as
\begin{multline}
    \ddx{\vect{x}_n}h(\vect{z},\vect{x}_i,\vect{x}_j) = \\
    \begin{cases}
        0  &\text{for } n=i \wedge n=j\\
        \left(\frac{2\vect{d}_{nj} + 2 \vect{z}}{-\sigma_z^2}\right)h(\vect{z},\vect{x}_n,\vect{x}_j)  &\text{for } n=i \wedge n \neq j \\
        -\left(\frac{2\vect{d}_{in} + 2 \vect{z}}{-\sigma_z^2}\right) h(\vect{z},\vect{x}_i,\vect{x}_n) &\text{for } n \neq i \wedge n = j \\
        0  &\text{for } n \neq i \wedge n \neq j 
    \end{cases} \,. \label{eq:partial_h}
\end{multline}
Considering we have a summation over $i$ and a summation over $j$ for every partial derivation toward $x_j$, we can identify 4 cases, two of which are trivial, where $x_n$ does not appear or $x_i-x_j=x_n-x_n=0$, the other two are very similar and can reuse partial results from the function evaluation, reducing the computational complexity.

\section{Optimisation Algorithm}
\label{ap:algorithm}

\begin{algorithm}[!h]
\caption{Bit labelling}\label{alg:bit_label}
\begin{algorithmic}[1]
\State $x \gets \Re^{M\times 2N}$ \Comment{\textit{Constellation}}
\State $m \gets [m_1,\ldots,m_{2N}]$ \Comment{\textit{Number of bits per dimension}}
\Procedure{AssignLabels}{$x,m$} \Comment{\textit{Map $m$ bits to $x$}}
    \State{$i\gets \textsc{Argsort}(x[:,1])$}
    \State{$D \gets \textsc{Ndim}(x)$}
    \If{$D=1$}%
        \State{$l(i) \gets \textsc{Graymap}(m)$}
        \LineComment{Assign Gray map to 1D constellation.}
    \Else
        
        \State{$l_{\mathrm{dim}} \gets \textsc{Graymap}(m[1])$} 
        \LineComment{Labels for first dimension.}
        \For{$j \gets 1,\ldots 2^{m[1]}$}
            \LineComment{\parbox{16em}{Recursively label the other dimensions for each label seperately.}}
            \State{$M_j \gets \prod_{d=2}^{D}2^{m_d}$} 
            \LineComment{Number labels in next dimensions.}
            \State{$s \gets i[M_j (j-1)+ (1,\ldots,M_j)]$}
            \LineComment{Selection for this label.}
            \State{$l_s \gets \textsc{AssignLabels}(x[s,2,\ldots,D],m[2,\ldots,D])$} 
            \LineComment{Labels for the selection.}
            \State{$l[s] \gets M_j l_{\mathrm{dim}}[j]+l_s$} \LineComment{Assign combination of labels.}
        \EndFor
    \EndIf\\
    \Return{$l$}
\EndProcedure
\end{algorithmic}
\end{algorithm}

\begin{figure}[h]
    \centering
    \begin{tikzpicture}[node distance=1.5cm, font=\footnotesize]
    \node (in1) [io,align=center] {Generate $x$\\ Assign labels};
    \node (pro1) [process, below of=in1, align=center] {$f(x)$ and $\nabla f(x)$};
    \node (pro2) [process, below of=pro1, align=center] {Find trust-region\\ step $s_k$};
    \node (pro3) [process, below of=pro2] {$f(x+s_k)$ and $\nabla f(x+s_k)$};
    \node (dec1) [decision, below of=pro3, align=center] {Update $x$, \\Hessian and\\ Trust-Region };
    \node (out1) [io, below of=dec1] {Save $x$};

        \draw [arrow,<-] (in1) -- ++(-2cm,0)  node[draw,circle,anchor=east,fill=black] {};
    \draw [arrow] (out1) -- ++(-2cm,0)  node[draw,circle,anchor=east,inner sep=2pt] {\begin{tikzpicture}
        \node[draw,circle,fill=black,inner sep=0, minimum width=2pt] {};
    \end{tikzpicture}};

    \draw [arrow] (in1) -- (pro1);
    \draw [arrow] (pro1) -- coordinate (tap1) (pro2);
    \draw [arrow] (pro2) -- (pro3);
    \draw [arrow] (pro3) -- coordinate (tap2) (dec1);

    \draw [arrow] (dec1) -- node[anchor=east] {Trust region\textless target} (out1);
    \draw [arrow] (dec1) -- node[anchor=north,align=center] {Trust region\\ \textgreater target} ++(3cm,0) |- (pro2);
    
    \draw [arrow, dotted] (tap1) -- +(-2cm,0) node[anchor=east, yshift=0cm, solid,-] {\begin{tikzpicture}
        \begin{axis}[grid=both,unit vector ratio=1 1,width =4cm,/tikz/font=\small,
                    xmin=-0.5,xmax=2.5,ymin=-0.5,ymax=2.5]
         \addplot [Set1-A, thick, only marks, mark=*, mark options={solid}, mark size=1pt,  quiver={u=-4*\thisrowno{2},v=-4*\thisrowno{3}}, -stealth ]
           table[x index=0,y index=1]{data/step0.txt};
        \end{axis}
        \end{tikzpicture}};
        
    \draw [arrow, dotted] (tap2) -- +(-2cm,0) node[anchor=east, yshift=0cm, solid,-] {\begin{tikzpicture}
        \begin{axis}[grid=both,unit vector ratio=1 1,width =4cm,/tikz/font=\small,
                    xmin=-0.5,xmax=2.5,ymin=-0.5,ymax=2.5]
        \addplot [Set1-A, thick, only marks, mark=*, mark options={solid},fill opacity=0.3,draw opacity=0, mark size=1pt, -stealth ]
           table[x index=0,y index=1]{data/step0.txt};
         \addplot [Set1-B, thick, only marks, mark=*, mark options={solid}, mark size=1pt,  quiver={u=-4*\thisrowno{6},v=-4*\thisrowno{7}}, -stealth ]
           table[x index=4,y index=5]{data/step0.txt};
        \end{axis}
        \end{tikzpicture}};
     
    \end{tikzpicture}     \caption{Process workflow of the optimisation algorithm for constellation $x$. In this process we used Eq.~\eqref{eq:gmi_function} in Eq.~\eqref{eq:chain_optimisation} for $f(x)$ and Eq.~\eqref{eq:chain_jacobian} for $\nabla f(x)$. The insets show a single quadrant of the constellation and its gradient for $x$ and $x+s_k$ of the first step.}
    \label{fig:optimisation_alg}
\end{figure}
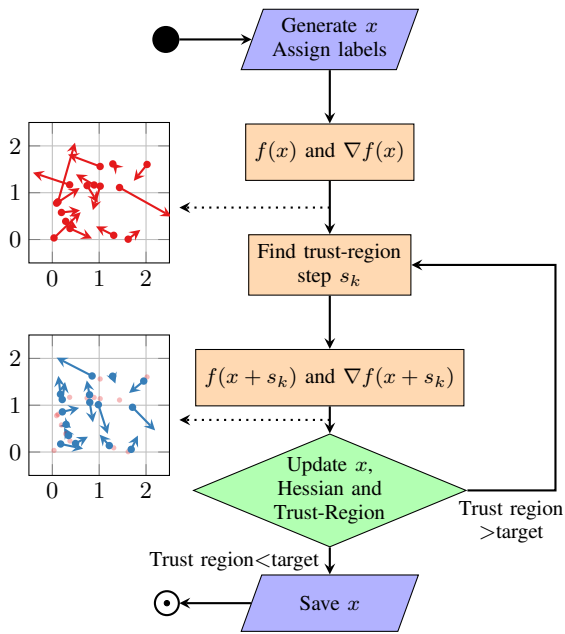

The optimisation algorithm consists of two parts, first the initial constellation and second the trust-region optimisation. We found that generating a few constellations as a starting point, optimising them with the trust-region algorithm until the selected stopping condition is reached, and keeping only the best performing constellation gave good results.

For the initial constellation, there are a few options. First, there is the use of the known `good'  constellations and binary labellings. Herein we have used square QAM and constellations proposed in \cite{Meric2015Approaching}. Alternatively, a lattice or a randomly generated constellation using a Gaussian prior as starting points were also used in this work. When using randomly assigned bit labels the optimisation algorithm converged to local minima, we only found good performance when bit labels were added. To obtain the bit labels, we have used the algorithm~\ref{alg:bit_label}, for which we need the following definitions: If $L$ is a list, then $\textsc{Argsort}(L)$ is the lexicographically least permutation of the indices $[0, \ldots , \textsc{Len}(L) - 1]$ such that $L[\textsc{Argsort}(L)]$ is sorted.
The procedure $\textsc{Graymap}(m)$ returns the indices for a 1D Gray-coded constellation, e.g., $\textsc{Graymap}(2) = [0,1,3,2]$ and $\textsc{Graymap}(3) = [0,1,3,2,6,7,5,4]$.

When the bit labelling is obtained for the Cartesian coordinates of the constellation, the bit labels resemble the Gray-coded \ac{QAM}. If fact, labelling a regular square constellation results in Gray-coded regular square \ac{QAM}. Similarly, when the bit labelling is obtained for the $N$-dimensional spherical coordinates, the constellations resemble AS-PSK or the constellations proposed in \cite{Meric2015Approaching}. We found that for lower SNR relative to the cardinality of the constellation, the latter performs better and for higher SNR the Cartesian basis for generating the bit labels works best.

The process of the trust-region optimisation is shown in Fig.~\ref{fig:optimisation_alg}. After generating a constellation and assigning bit labels to it, a trust-region algorithm is applied \cite{Nocedal2006Numerical}. The algorithms start with calculating the function value and its gradient. The inverse Hessian $B$ is assumed to be an identity matrix and the trust region to be of size 1. Steihaug's conjugate gradient method\cite{Steihaug1983Conjugate} is used to find trust-region step $s_k$ and the function and gradient at $x+s_k$ are evaluated. The ratio between the actual reduction and the predicted reduction \cite[Eq.~(4.4)]{Nocedal2006Numerical} is used to update the constellation and the trust region size. If it is greater than 0, the new $x$ will be $x+s_k$, otherwise $s_k$ is discarded. Note that the actual reduction can be negative and move the constellation away from its current local minimum. The size of the trust region is reduced or increased based on whether the ratio is below $0.2$ or above $0.8$ respectively. When the trust-region size has shrunk smaller than a set target (we used $3\times10^{-4}$) the optimisation is considered complete. 

\section*{Acknowledgment} %
The authors are grateful to Dr Bin Chen for discussions on the topic of geometric constellation shaping and for sharing the coordinates of the OS-128 constellation. %

\bibliographystyle{IEEEtran}

\begin{thebibliography}{10}
\providecommand{\url}[1]{#1}
\csname url@samestyle\endcsname
\providecommand{\newblock}{\relax}
\providecommand{\bibinfo}[2]{#2}
\providecommand{\BIBentrySTDinterwordspacing}{\spaceskip=0pt\relax}
\providecommand{\BIBentryALTinterwordstretchfactor}{4}
\providecommand{\BIBentryALTinterwordspacing}{\spaceskip=\fontdimen2\font plus
\BIBentryALTinterwordstretchfactor\fontdimen3\font minus
  \fontdimen4\font\relax}
\providecommand{\BIBforeignlanguage}[2]{{%
\expandafter\ifx\csname l@#1\endcsname\relax
\typeout{** WARNING: IEEEtran.bst: No hyphenation pattern has been}%
\typeout{** loaded for the language `#1'. Using the pattern for}%
\typeout{** the default language instead.}%
\else
\language=\csname l@#1\endcsname
\fi
#2}}
\providecommand{\BIBdecl}{\relax}
\BIBdecl

\bibitem{Ghazisaeidi2016Transoceanic}
A.~{Ghazisaeidi}, I.~F. d.~{Jauregui Ruiz}, R.~{Rios-Muller}, L.~{Schmalen},
  P.~{Tran}, P.~{Brindel}, A.~C. {Meseguer}, Q.~{Hu}, F.~{Buchali},
  G.~{Charlet}, and J.~{Renaudier}, ``{65Tb/s} transoceanic transmission using
  probabilistically-shaped {PDM-64QAM},'' in \emph{European Conference on
  Optical Communication (ECOC) - Post Deadline Paper}, 2016.

\bibitem{Olsson2018Probabilistically}
S.~L. Olsson, J.~Cho, S.~Chandrasekhar, X.~Chen, P.~J. Winzer, and S.~Makovejs,
  ``Probabilistically shaped pdm 4096-qam transmission over up to 200 km of
  fiber using standard intradyne detection,'' \emph{Opt. Express}, vol.~26,
  no.~4, pp. 4522--4530, Feb 2018.

\bibitem{Ionescu2019Transmission}
M.~{Ionescu}, D.~{Lavery}, A.~{Edwards}, E.~{Sillekens}, L.~{Galdino},
  D.~{Semrau}, R.~I. {Killey}, W.~{Pelouch}, S.~{Barnes}, and P.~{Bayvel},
  ``74.38 {Tb/s}0 transmission over 6300 km single mode fiber with hybrid
  {EDFA/Raman} amplifiers,'' in \emph{2019 Optical Fiber Communications
  Conference and Exhibition (OFC)}, 2019.

\bibitem{Galdino2020Optical}
L.~{Galdino}, A.~{Edwards}, W.~{Yi}, E.~{Sillekens}, Y.~{Wakayama},
  T.~{Gerard}, W.~S. {Pelouch}, S.~{Barnes}, T.~{Tsuritani}, R.~I. {Killey},
  D.~{Lavery}, and P.~{Bayvel}, ``Optical fibre capacity optimisation via
  continuous bandwidth amplification and geometric shaping,'' \emph{IEEE
  Photonics Technology Letters}, vol.~32, no.~17, pp. 1021--1024, 2020.

\bibitem{Wakayama2019Increasing}
Y.~{Wakayama}, E.~{Sillekens}, L.~{Galdino}, D.~{Lavery}, R.~I. {Killey}, and
  P.~{Bayvel}, ``Increasing achievable information rates with pilot-based {DSP}
  in standard intradyne detection,'' in \emph{European Conference on Optical
  Communication (ECOC)}, 2019.

\bibitem{Cai2018Transmission}
J.~{Cai}, H.~G. {Batshon}, M.~V. {Mazurczyk}, O.~V. {Sinkin}, D.~{Wang},
  M.~{Paskov}, W.~W. {Patterson}, C.~R. {Davidson}, P.~C. {Corbett}, G.~M.
  {Wolter}, T.~E. {Hammon}, M.~A. {Bolshtyansky}, D.~G. {Foursa}, and A.~N.
  {Pilipetskii}, ``70.46 {Tb/s} over 7,600 km and 71.65 {Tb/s} over 6,970 km
  transmission in {C+L} band using coded modulation with hybrid constellation
  shaping and nonlinearity compensation,'' \emph{Journal of Lightwave
  Technology}, vol.~36, no.~1, pp. 114--121, 2018.

\bibitem{Jones2018Deep}
R.~T. {Jones}, T.~A. {Eriksson}, M.~P. {Yankov}, and D.~{Zibar}, ``Deep
  learning of geometric constellation shaping including fiber nonlinearities,''
  in \emph{European Conference on Optical Communication (ECOC)}, 2018.

\bibitem{Essiambre2020Increasing}
R.-J. Essiambre, R.~Ryf, M.~Kodialam, B.~Chen, M.~Mazur, J.~Bonetti,
  R.~Veronese, H.~Huang, A.~Gupta, F.~A. Aoudia, E.~Burrows, D.~Grosz,
  L.~Palmieri, X.~Chen, N.~K. Fontaine, H.~Chen, and M.~Sellathurai,
  ``Increased reach of long-haul transmission using a constant-power 4d format
  designed using neural networks,'' in \emph{European Conference on Optical
  Communication (ECOC)}, 2020, pp. Mo1E--5.

\bibitem{Veeru2020End}
V.~T. abd Toshiaki Koike-Akino, Y.~Wang, D.~Millar, K.~Kojima, and K.~Parsons,
  ``End-to-end deep learning for phase noise-robust multi-dimensional geometric
  shaping,'' in \emph{European Conference on Optical Communication (ECOC)},
  2020, pp. Th1D--4.

\bibitem{Fehenberger2016Probabilistic}
T.~{Fehenberger}, A.~{Alvarado}, G.~{Böcherer}, and N.~{Hanik}, ``On
  probabilistic shaping of quadrature amplitude modulation for the nonlinear
  fiber channel,'' \emph{Journal of Lightwave Technology}, vol.~34, no.~21, pp.
  5063--5073, 2016.

\bibitem{Geller2016Shaping}
O.~{Geller}, R.~{Dar}, M.~{Feder}, and M.~{Shtaif}, ``A shaping algorithm for
  mitigating inter-channel nonlinear phase-noise in nonlinear fiber systems,''
  \emph{Journal of Lightwave Technology}, vol.~34, no.~16, pp. 3884--3889,
  2016.

\bibitem{Renner2017Experimental}
J.~{Renner}, T.~{Fehenberger}, M.~P. {Yankov}, F.~{Da Ros}, S.~{Forchhammer},
  G.~{Böcherer}, and N.~{Hanik}, ``Experimental comparison of probabilistic
  shaping methods for unrepeated fiber transmission,'' \emph{Journal of
  Lightwave Technology}, vol.~35, no.~22, pp. 4871--4879, 2017.

\bibitem{Sillekens2018Simple}
E.~Sillekens, D.~Semrau, G.~Liga, N.~A. Shevchenko, Z.~Li, A.~Alvarado,
  P.~Bayvel, R.~I. Killey, and D.~Lavery, ``A simple nonlinearity-tailored
  probabilistic shaping distribution for square {QAM},'' in \emph{Optical Fiber
  Communication Conference}.\hskip 1em plus 0.5em minus 0.4em\relax Optical
  Society of America, 2018, p. M3C.4.

\bibitem{Zhang2017Design}
S.~Zhang, F.~Yaman, E.~Mateo, T.~Inoue, K.~Nakamura, and Y.~Inada, ``Design and
  performance evaluation of a gmi-optimized 32qam,'' in \emph{European
  Conference on Optical Communication (ECOC)}, 2017.

\bibitem{Sillekens2018Experimental}
E.~{Sillekens}, D.~{Semrau}, D.~{Lavery}, P.~{Bayvel}, and R.~I. {Killey},
  ``Experimental demonstration of geometrically-shaped constellations tailored
  to the nonlinear fibre channel,'' in \emph{European Conference on Optical
  Communication (ECOC)}, 2018.

\bibitem{chen2020analysis}
B.~Chen, A.~Alvarado, S.~van~der Heide, M.~van~den Hout, H.~Hafermann, and
  C.~Okonkwo, ``Analysis and experimental demonstration of orthant-symmetric
  four-dimensional 7 bit/{4D-Sym} modulation for optical fiber communication,''
  pp. 2737--2753, 2021.

\bibitem{Dzieciol2020Geometric}
H.~{Dzieciol}, G.~{Liga}, E.~{Sillekens}, P.~{Bayvel}, and D.~{Lavery},
  ``Geometric shaping of 2-dimensional constellations in the presence of laser
  phase noise,'' \emph{Journal of Lightwave Technology}, p. Preprint, 2020.

\bibitem{Beda1959Pfa}
L.~M. Beda, L.~N. Korolev, N.~V. Sukkikh, and T.~S. Frolova, ``Programs for
  automatic differentiation for the machine {BESM},'' Institute for Precise
  Mechanics and Computation Techniques, Academy of Science, Moscow, USSR,
  {T}echnical {R}eport, 1959, (In Russian).

\bibitem{Rall1981Automatic}
L.~B. Rall, Ed., \emph{Automatic Differentiation: Techniques and
  Applications}.\hskip 1em plus 0.5em minus 0.4em\relax Springer Berlin
  Heidelberg, 1981.

\bibitem{Foschini1974Optimization}
G.~Foschini, R.~Gitlin, and S.~Weinstein, ``Optimization of two-dimensional
  signal constellations in the presence of gaussian noise,'' \emph{IEEE
  Transactions on Communications}, vol.~22, no.~1, pp. 28--38, 1974.

\bibitem{Golub1996Matrix}
G.~H. Golub and C.~F. Van~Loan, \emph{Matrix Computations}, 3rd~ed.\hskip 1em
  plus 0.5em minus 0.4em\relax Johns Hopkins, 1996.

\bibitem{Splett1993Ultimate}
A.~Splett, C.~Kurtzke, and K.~Petermann, ``Ultimate transmission capacity of
  amplified optical fiber communication systems taking into account fiber
  nonlinearities,'' in \emph{1993 The European Conference on Optical
  Communication (ECOC)}, 1993.

\bibitem{Poggiolini2012GN}
P.~Poggiolini, ``The gn model of non-linear propagation in uncompensated
  coherent optical systems,'' \emph{Journal of Lightwave Technology}, vol.~30,
  no.~24, pp. 3857--3879, 2012.

\bibitem{Dar2013Properties}
R.~Dar, M.~Feder, A.~Mecozzi, and M.~Shtaif, ``Properties of nonlinear noise in
  long, dispersion-uncompensated fiber links,'' \emph{Opt. Express}, vol.~21,
  no.~22, pp. 25\,685--25\,699, Nov 2013.

\bibitem{dar2014on}
R.~Dar \emph{et~al.}, ``On shaping gain in the nonlinear fiber-optic channel,''
  \emph{IEEE International Symposium on Information Theory}, pp. 2794--2798,
  June 2014.

\bibitem{Semrau2019AClosedForm}
D.~Semrau, R.~I. Killey, and P.~Bayvel, ``A closed-form approximation of the
  gaussian noise model in the presence of inter-channel stimulated raman
  scattering,'' \emph{J. Lightwave Technol.}, vol.~37, no.~9, pp. 1924--1936,
  May 2019.

\bibitem{Semrau2019AModulation}
D.~Semrau, E.~Sillekens, R.~I. Killey, and P.~Bayvel, ``A modulation format
  correction formula for the gaussian noise model in the presence of
  inter-channel stimulated {Raman} scattering,'' \emph{J. Lightwave Technol.},
  vol.~37, no.~19, pp. 5122--5131, Oct 2019.

\bibitem{Shannon1948Mathematical}
C.~E. Shannon, ``A mathematical theory of communication,'' \emph{Bell System
  Technical Journal}, vol.~27, no.~3, pp. 379--423, 1948.

\bibitem{Cover2005Elements}
\BIBentryALTinterwordspacing
T.~M. Cover and J.~A. Thomas, \emph{Elements of Information Theory},
  2nd~ed.\hskip 1em plus 0.5em minus 0.4em\relax Wiley, Apr. 2005. [Online].
  Available: \url{https://doi.org/10.1002/047174882x}
\BIBentrySTDinterwordspacing

\bibitem{Caire1988Bit-interleaved}
G.~{Caire}, G.~{Taricco}, and E.~{Biglieri}, ``Bit-interleaved coded
  modulation,'' \emph{IEEE Transactions on Information Theory}, vol.~44, no.~3,
  pp. 927--946, 1998.

\bibitem{Alvarado2018Achievable}
A.~Alvarado, T.~Fehenberger, B.~Chen, and F.~M.~J. Willems, ``Achievable
  information rates for fiber optics: Applications and computations,'' \emph{J.
  Lightwave Technol.}, vol.~36, no.~2, pp. 424--439, Jan 2018.

\bibitem{Broyden1970Convergence}
C.~G. Broyden, ``The convergence of a class of double-rank minimization
  algorithms 1. general considerations,'' \emph{{IMA} Journal of Applied
  Mathematics}, vol.~6, no.~1, pp. 76--90, 1970.

\bibitem{Fletcher1970Approach}
R.~Fletcher, ``A new approach to variable metric algorithms,'' \emph{The
  Computer Journal}, vol.~13, no.~3, pp. 317--322, Mar. 1970.

\bibitem{Goldfarb1970Family}
D.~Goldfarb, ``A family of variable-metric methods derived by variational
  means,'' \emph{Mathematics of Computation}, vol.~24, no. 109, pp. 23--23,
  Jan. 1970.

\bibitem{Shanno1970Conditioning}
D.~F. Shanno, ``Conditioning of quasi-{Newton} methods for function
  minimization,'' \emph{Mathematics of Computation}, vol.~24, no. 111, pp.
  647--647, Sep. 1970.

\bibitem{Coleman1996}
T.~F. Coleman and Y.~Li, ``An interior trust region approach for nonlinear
  minimization subject to bounds,'' \emph{{SIAM} Journal on Optimization},
  vol.~6, no.~2, pp. 418--445, may 1996.

\bibitem{Byrd1987Trust}
R.~H. Byrd, R.~B. Schnabel, and G.~A. Shultz, ``A trust region algorithm for
  nonlinearly constrained optimization,'' \emph{SIAM Journal on Numerical
  Analysis}, vol.~24, no.~5, pp. 1152--1170, 1987.

\bibitem{Byrd1996Analysis}
R.~H. Byrd, H.~F. Khalfan, and R.~B. Schnabel, ``Analysis of a symmetric
  rank-one trust region method,'' \emph{{SIAM} Journal on Optimization},
  vol.~6, no.~4, pp. 1025--1039, nov 1996.

\bibitem{Conn1991Convergence}
A.~R. Conn, N.~I.~M. Gould, and P.~L. Toint, ``Convergence of quasi-newton
  matrices generated by the symmetric rank one update,'' \emph{Mathematical
  Programming}, vol.~50, no.~1, pp. 177--195, Mar 1991.

\bibitem{Nocedal2006Numerical}
J.~Nocedal and S.~Wright, \emph{Numerical Optimization}, 2nd~ed.\hskip 1em plus
  0.5em minus 0.4em\relax Springer New York, 2006.

\bibitem{Steihaug1983Conjugate}
T.~Steihaug, ``The conjugate gradient method and trust regions in large scale
  optimization,'' \emph{SIAM Journal on Numerical Analysis}, vol.~20, no.~3,
  pp. 626--637, 1983.

\bibitem{Liga2020Extending}
G.~Liga, A.~Barreiro, H.~Rabbani, and A.~Alvarado, ``Extending fibre nonlinear
  interference power modelling to account for general dual-polarisation 4d
  modulation formats,'' \emph{Entropy}, vol.~22, no.~11, 2020.

\bibitem{Meric2015Approaching}
H.~{Méric}, ``Approaching the gaussian channel capacity with {APSK}
  constellations,'' \emph{IEEE Communications Letters}, vol.~19, no.~7, pp.
  1125--1128, July 2015.

\bibitem{Sun1993Approaching}
{Feng-Wen Sun} and H.~C.~A. {van Tilborg}, ``Approaching capacity by
  equiprobable signaling on the gaussian channel,'' \emph{IEEE Transactions on
  Information Theory}, vol.~39, no.~5, pp. 1714--1716, Sep. 1993.

\bibitem{Forney1984Effecient}
G.~Forney, R.~Gallager, G.~Lang, F.~Longstaff, and S.~Qureshi, ``Efficient
  modulation for band-limited channels,'' \emph{IEEE Journal on Selected Areas
  in Communications}, vol.~2, no.~5, pp. 632--647, 1984.

\bibitem{Menyuk1987Nonlinear}
C.~Menyuk, ``Nonlinear pulse propagation in birefringent optical fibers,''
  \emph{IEEE Journal of Quantum Electronics}, vol.~23, no.~2, pp. 174--176,
  1987.

\bibitem{Standard2014DVBS2X}
\emph{Digital Video Broadcasting {(DVB)}; Second generation framing structure,
  channel coding and modulation systems for Broadcasting, Interactive Services,
  News Gathering and other broadband satellite applications; Part 2: {DBV-S2}
  Extensions {(DVB-S2X)}}, European Standard ETSI EN 302 307-2, 2014.

\bibitem{Abramowitz1965Handbook}
M.~Abramowitz and I.~A. Stegun, \emph{\BIBforeignlanguage{eng}{Handbook of
  mathematical functions with formulas, graphs and mathematical tables}},
  9th~ed.\hskip 1em plus 0.5em minus 0.4em\relax New York: Dover, 1965.

\end{thebibliography}
\end{document}